# Output fusion of MPC and PID and its application in intelligent layered water injection of oilfield

*Yuan-Long Yue [a]*, Hao-Yang Wen [a], Xin Zuo [a], Mao Sheng [a], Fu-Chao Sun [b]*

*[a] China University of Petroleum (Beijing), Beijing, 102249, China*

*[b] Research institute of petroleum exploration & development, Beijing, 100083, China*

**Abstract:** In the area of intelligent layered water injection in the oilfield, wave code communication based on pipe column medium as a wireless data transmission method between the wellhead monitoring system and underground water distributor is the current research hotspot. However, the existing wave code communication technology is time-consuming and has a high symbol error rate, especially in the low permeability reservoirs communication process also suffers from the problem of small wave code limit amplitude that leads to a low recognition rate. The fundamental reason is that the existing method of controlling the generation of wave codes cannot quickly drive the regular fluctuation of the fluid in the tube column, and the generated wave codes cannot meet the decoding requirements, resulting in low communication efficiency. To improve the dynamic response performance of wave code communication, this paper proposes an output optimal fusion control method based on MPC-PID. Firstly, depending on the well structure and the flow-pressure characteristics of the layer, the steady-state model between the differential pressure and flow of the whole well and different layer sections is established for layered water injection, and the corresponding wave code amplitude at the steady-state operating point of different layer sections is solved, the numerical calculation verifies that the increase of the nozzle opening in a single layer section will drive the pressure and flow curve of the whole well downward. Secondly, combining the dynamic response characteristics and steady-state model of the whole-well water distribution equipment, a dynamic model of layered intelligent water injection is established, and the generation process of the wave code is defined; Finally, the MPC-PID optimal fusion control algorithm structure is designed to derive the fusion control law that minimizes the cost function under fixed weights, , and the optimal weights are calculated by combining the internal model structure of controller, so the optimization performance of each algorithm in the optimal fusion control is balanced. By analyzing the control simulation results, the fast response characteristics of the fusion control method are verified. Meanwhile, the simulation comparison experiments of fast wave code communication under different methods are conducted with the actual working conditions, the results show that the fusion control method has both fast tracking control capability and strong robustness, which effectively enhances the efficiency of wave code communication and shortens the wave code operation time.

# 1 Introduction

The supply conflict of petroleum energy in today's world is becoming increasingly prominent, and how to improve the recovery ratio with the existing proven reserves is becoming an issue of growing importance (Song et al., 2021). Artificial oil drive technology is an effective means to increase production in oilfield production, and the existing efficient oil drive methods mainly include carbon dioxide drive, chemical drive, nano flooding, water drive (Li et al., 2021; Liang et al.,2021; Mohamed., et al), but water drive is still the most widely used technology due to the limitation of cost and research level (Ajoma et al., 2020; Jiang et al.,2019; Wang et al., 2018). However, the normal water drive method does not apply to current tight reservoirs with a high reserve share and wide geographic distribution (Luo et al., 2017; Guo et al., 2021), because of the high opening pressure of such reservoirs (Wang et al., 2008), complex geological conditions, and obvious differences in physical properties such as pore structure and denseness of rocks in each layer section (Zhao et al., 2021), so it is harder to maintain a full well water injection qualification rate using the traditional water drive method, which leads to unrealized single well target production and thus lower economic efficiency (Zhou et al., 2021). The traditional water injection method uses generalized co-injection and uses armor or steel pipe cables for communication, with low extraction quality and safety risks during testing. In contrast, in single-well multi-layer segment injection operations, for layers with a certain permeability, when the flow does not reach optimal values, it leads to insufficient pressure and reduced recovery, which indirectly reduces the efficiency of energy use. However, oversized flow can lead to layer pressure overload, harming the geological structure of the layer segment, causing single layer bursts and ineffective water circulation, and a dramatic increase in water content (Ganat et al., 2015), so it is crucial to keep the flow values of the layer segment matching the optimal value (Jing et al., 2017). To this end, the layered segment fine water injection technology was developed, which independently sets mining parameters depending on the geological characteristics of different layers, effectively solving the problems of inter-layer interference and uneven flow advance (Zhang et al., 2020), thus realizing the optimal control of pressure and flow in the layers and ensuring the recovery rate of the layers with different characteristics (Jia et al., 2020).

In layered water injection operations, the reservoirs are buried deeper in the ground and the injection status parameters are more difficult to obtain, making it impossible to achieve real-time monitoring of the underground water injection status. Therefore, it is crucial to realize the transmission communication between the wellhead and the well. Existing technologies use interoperable cables for communication. In 2019, Zhou applied the layered water injection technology based on preset cables experimentally in Daqing oilfield (Zhou, 2019), which effectively reduced the cost of manual well logging, realized real-time monitoring of downhole data, and improved the qualification rate of layered water injection. However, it is obvious that this technique of cable repair is inconvenient and the maintenance cost is high. (Li, 2019). In 2020 Tong studied the wireless intelligent water injection technology based on a computer network (Tong, 2020), which uses PID algorithm to control the water distribution flow, compared with cable transmission whose advantage is that it does not require larger upfront investment and cable maintenance, the disadvantage is that wireless communication is not timely, more dependent on network communication, while the control algorithm is simple and cannot meet the dynamic performance requirements of flow control in different layer sections. In 2020, a wave code communication technology was proposed (Yao et al., 2020), which decodes and converts fluid fluctuation signals into control signals to achieve this. This technology uses fluid as the signal medium, which effectively solves the problems of difficult protection and the high operating cost of traditional intelligent layered water injection communication equipment. However, the use of simple control algorithms for flow regulation generates many wave code transmission links and long delays, resulting in slow pressure response and high symbol error rates, which cannot meet the needs of efficient decoding, fast following of target values, and real-time communication, resulting in power wastage and actuator wear. All in all, the study of advanced control algorithms for intelligent water injection control of complex layer segments has high economic efficiency and applicability.

Layered water injection is a typical multivariate coupling process, which requires the design of targeted control methods. Model predictive control (MPC), as a commonly used advanced control algorithm, has certain inherent decoupling capabilities, and its control law is based on the characteristics of the model, making its control specialization better than PID control algorithms. However, for variable objects that require fast response such as flow, the response speed, and anti-interference characteristics will not meet the requirements due to their calculation speed (Qian et al.,2021). In addition, the model predictive control algorithm is more complex and the parameters are more inconvenient to adjust, so it is generally used as a master controller in conjunction with proportional-integral control. Therefore, Ravendra proposed an MPC-PID cascade control algorithm for the pharmaceutical process (Ravendra et al., 2013) and performed process simulation of direct tablet pressing process using gPROMS software. The results showed that the product qualification rate was significantly improved compared to the string-level PID control algorithm and the MPC algorithm alone (Ravendra et al., 2014). Its use of MPC output to do the main loop set value, PID controller as the lower-level controller determines the main dynamic performance of the system, compared to the PID control quality improved. However, the output overshoot is still large, the oscillation cannot be eliminated. Meanwhile, the direct introduction of PID will weaken the robustness of MPC, in the case of large model gain coefficient and mismatch will seriously deteriorate the system performance.

To improve the dynamic response performance of control-generated wave codes, this paper proposes a multivariate MPC-PID-based output-weighted fusion control algorithm for driving control fluid fluctuations to quickly generate high-quality wave codes, thus enabling efficient communication between the wellhead and underground water distributors. Fusion control integrates the advantages of various advanced control algorithms and changes the weights according to the actual needs of control. In 2013, Yuan proposed the integral-fusion-Smith fusion control algorithm (Yuan, 2013), which overcomes the weaknesses of the system with large time lag and difficulty in control when the model is mismatched, and its control accuracy is high and the dynamic performance index of the system is improved while showing better robustness. In 2020, Deng et al. proposed a weighted fusion algorithm of fuzzy PID control and self-anti-disturbance control (ADRC) for the optimal control of proton exchange membrane fuel cells (PEMFC) (Deng et al., 2020), the results show that the response overshoot and error become smaller, and the immunity of the system is greatly improved, solving the problem of insufficient oxygen during sudden load changes. In the proposed algorithm, MPC is placed in the outer loop of the algorithm, which improves the delay processing capability and robustness of the system, and the PID cascade link in the inner loop is responsible for improving the response speed and enhancing the ability to resist sudden disturbances, and the two are fused with optimal weights so that the optimization effectiveness of the dual algorithms can reach the optimal equilibrium state and achieve fast and accurate control.

In Section 2, a steady-state model is set up by analyzing the good structure and the relationship between differential pressure and flow in the whole well, which solves for the amplitude of the wave code. Then, a dynamic mathematical model of fluid fluctuation is established by combining the dynamic characteristics of water distribution equipment, completing the dynamic description of the wave code generation process. Section 3 derives the principle of the fusion control algorithm and solves the optimal control law under fixed weights, the error-free characteristics and closed-loop gain of multivariable control are also derived by combining the internal mode structure, the impact of weight optimization on system performance is quantitatively analyzed, and the optimization mechanism of wave code performance under the fusion control algorithm is defined. Section 4 is simulation. The results show that the fusion control algorithm improves the response speed, and solves the delay problem of wave codes in long-distance pipeline transmission under the condition that the common parameters of each algorithm are the same. To address the problem of high symbol error rate in wave code transmission, the output signal is made equivalent to a square wave by reducing the overshoot and the oscillation amplitude of the system, therefore achieving fast and accurate decoding of pressure waves. For the problems of inconvenient loading and complicated operation of MPC algorithm model, robustness simulation under discrete sampling anomaly and model mismatch state are designed. The proposed algorithm in the simulation demonstrates relatively good dynamic performance, improves the energy utilization efficiency, and achieves intelligent water injection controlling in the same well in the stratified section. Finally, conclusions are drawn. The above paper architecture and system control flow are shown in Fig. 1.

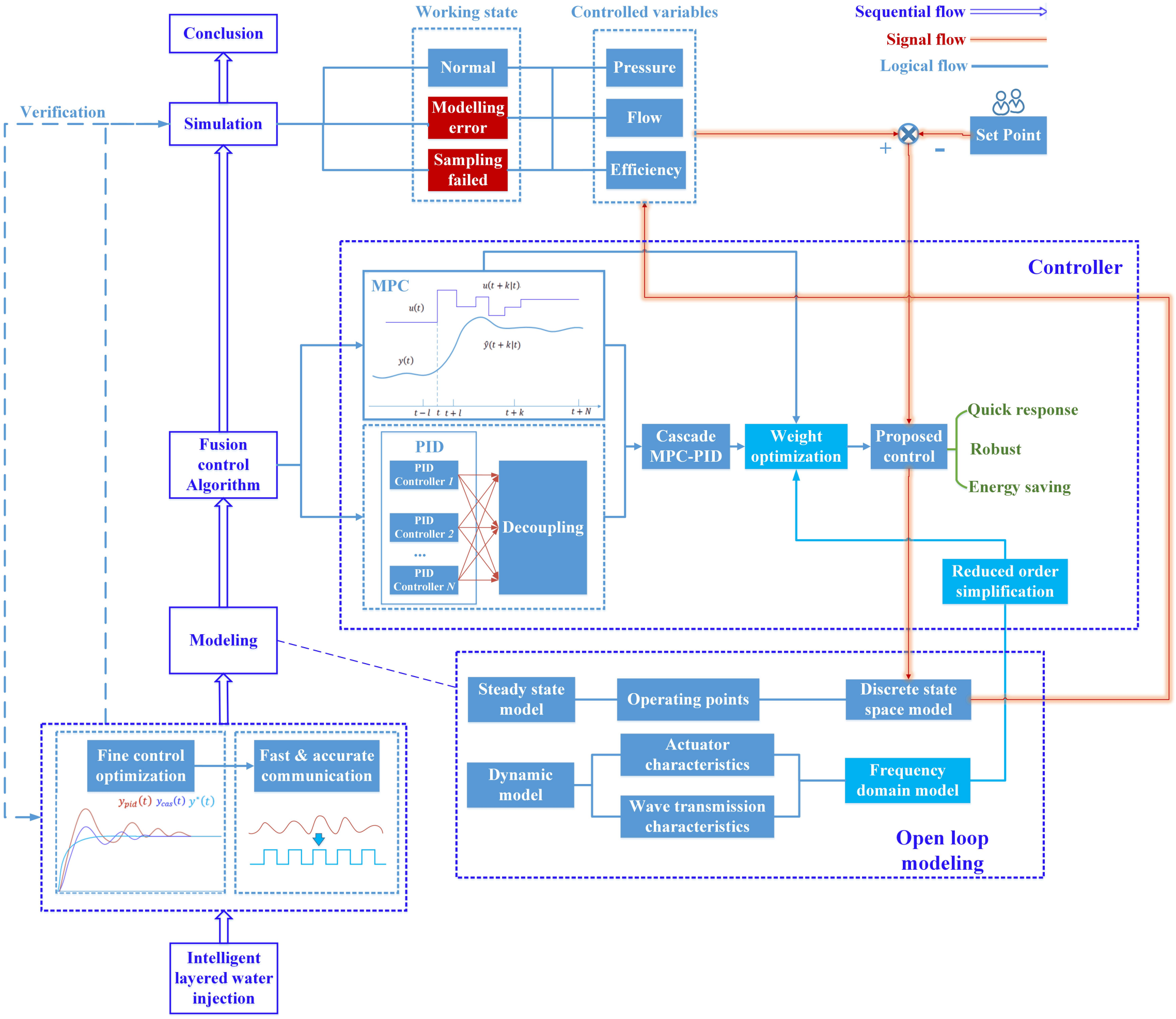

**Fig 1 Paper content and system control flow chart**

# 2 Layered water injection modeling

## 2.1 Steady-state model

In this section, a steady-state model of the layered intelligent water injection process is established by combining actual working conditions. A single-well layered water injection model is considered, consisting of N layers with different geological characteristics and a layered water injection control system, which includes ground control valves, water distributors with water nozzle groups set underground, water injection packer, pumping machines, and a central control system, etc. (Severiano et al., 2019). Accordingly, the overall architecture of the intelligent layered water injection system is shown in Fig. 2.

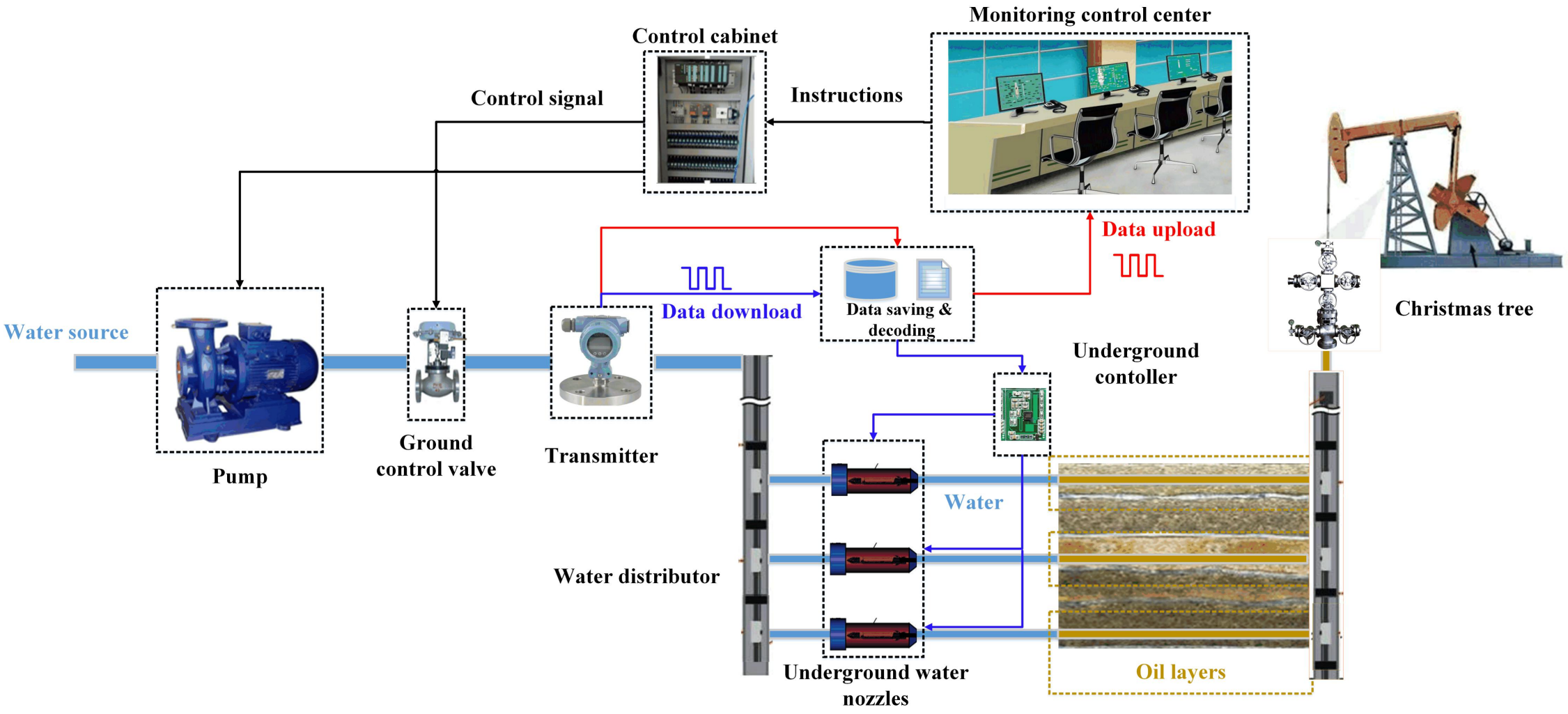

**Fig.2-Schematic diagram of the overall architecture of the intelligent layered water injection system**

As shown in the figure, in the intelligent injection distribution system, wave code communication between the wellhead and the underground distributor is carried out through periodic adjustment of the regulating valve. In the data upload stage, the underground water nozzle cyclically changes the opening degree and transmits the fluctuation information of relevant physical quantities to the database through sensors, which is retrieved and decoded by the monitoring center to obtain the downhole layered water injection parameters to judge the working status of the well accordingly. During the data transmission stage, the user sends instructions to control the ground valve group, corresponding to the generation of periodic fluctuation data information, which is decoded and transmitted to the underground controller loaded with control algorithms that drives the water distributor to finely adjust the spout opening degree for intelligent flow control. The user can set the parameters of the pump and controller in the ground control center during the operation. Combining the existing theoretical basis of wave code communication with the actual operating conditions, the operation process of intelligent layered water injection by wave code communication shown in Fig. 3 is defined.

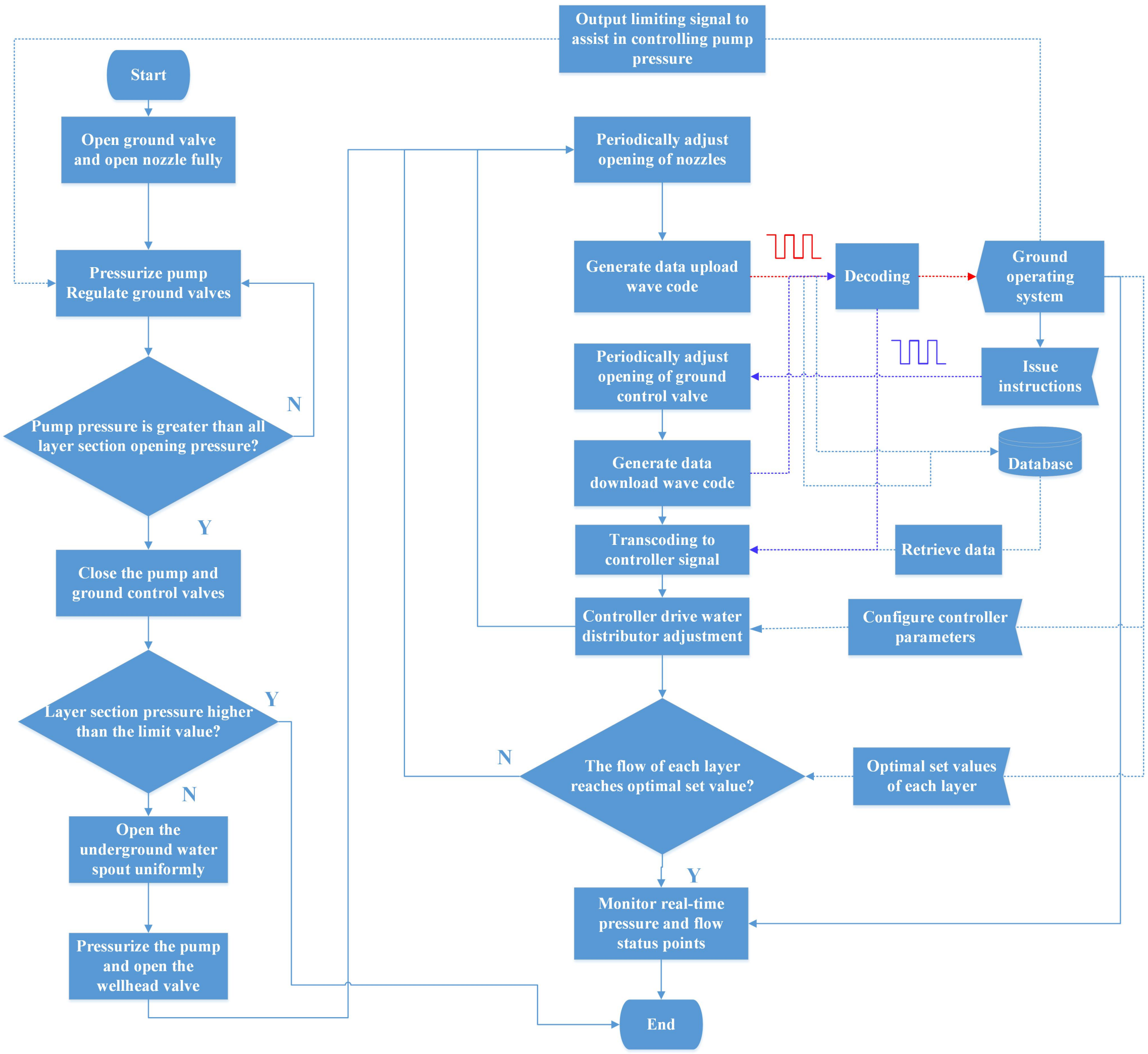

**Fig 3 Schematic diagram of layered water injection operation flow**

As shown in the figure, before the wave code communication control, the fracturing treatment of the layered injection well is first carried out, which can improve the permeability of layers to a certain extent and contribute to the production (Gao et al., 2022; Han et al., 2021). To improve the fracturing efficiency, the water distributor drives the downhole nozzle to be fully open and the ground control valve is set to open $C_0$. where the Nth layer characteristic is described using the following relationship:

$$p_{(N)} = k_{(N)} q_{v(N)} + b_{(N)} \tag{1}$$

$k$ is inversely proportional to the layer permeability and $b$ indicates the initial opening pressure of the layer. The flow is allowed to be injected into the layer when the external pressure is greater than this value (Zhang et al., 2013; Xiong et al., 2009). From the layer properties, it is known that full well fracturing can be achieved when the pump pressure is greater than the opening pressure of all layers. This leads to:

$$P_0 \geq b_{(t)max}, t = 1,2, \ldots, N \tag{2}$$

During the fracturing process, the pressure must not exceed the limit, otherwise, there will be water run-off to make the layer section penetrate. Set the safety pressure as $P_m$, and then from the differential pressure flow relationship of the throttling element can obtain the value range of the whole well flow value $Q_f$.

$$\sum_{i=1}^{N} C_0 \sqrt{b_{(t)max} - b_{(i)}} \leq Q_f < \sum_{i=1}^{N} C_0 \sqrt{P_m - b_{(i)}} \tag{3}$$

After the completion of the fracturing, the pipeline of each layer section is filled with fluid. Since the pump is closed, the flow values are zero and there is no flow in the pipeline, so the initial pressure of the layer will be transferred to the front of the nozzle through the fluid. Set $P$ as the pressure after the valve and in front of the nozzle, from which we can know the necessary conditions for the injection flow value of the whole well layer section as follow:

$$P \geq b_{(t)max} \tag{4}$$

The low limit of the full well flow at this point can be obtained as:

$$Q_{0l} = \sum_{i=1}^{N} \frac{C_{(i)}\sqrt{C_{(i)}{}^2 k_{(i)}{}^2 - 4\left(b_{(i)} - b_{(t)max}\right)} - C_{(i)}{}^2 k_{(i)}}{2} \tag{5}$$

Entering the water injection adjustment process and studying the equivalent steady-state modeling method between the incoming water at the wellhead and the layer section, it is necessary to specify the steady-state pressure-flow operating point under different equipment parameters. Set the $i$-th layer section nozzle opening from $C_{(i)0}$ to $C_{(i)}$, the $j$th layer flow steady state value becomes $q_{v(ij)}$, the downhole post-valve pressure is $P$, the incoming water pressure is $P_0$, and the ground control valve opening is $C_0$. Firstly, the differential pressure flow relationship of the wellhead section is listed as:

$$P = P_0 - \frac{\left(\sum_{j=1}^{N} q_{v(ij)} + q_{v(i)}\right)^2}{C_0{}^2}, \{j \neq i\} \tag{6}$$

Extending the above differential pressure flow relationships to the full well:

$$\begin{cases} q'_{v(i)} = \dfrac{C_{(i)}\sqrt{C_{(i)}{}^2 k_{(i)}{}^2 - 4\left(b_{(i)} - P\right)} - C_{(i)}{}^2 k_{(i)}}{2} \\ q'_{v(ij)} = \dfrac{C_{(j)}\sqrt{C_{(j)}{}^2 k_{(j)}{}^2 - 4\left(b_{(j)} - P\right)} - C_{(j)}{}^2 k_{(j)}}{2} \end{cases} \tag{7}$$

Flow values should be required to meet all layer opening and pressure limiting conditions:

$$\begin{cases} max\left(Q_{0l}, Q'_{0l}\right) \leq \sum_{i=1}^{N} q_{v(ij)} + q_{v(i)} < 2C_0\sqrt{P_m - b_{(t)min}} \\ q_{v(i)0}, q_{v(i)} < C_0\sqrt{P_m - b_{(i)}}, q_{v(ij)} < C_0\sqrt{P_m - b_{(j)}} \end{cases} \tag{8}$$

Where $C$ is the nozzle throttling coefficient:

$$C = \frac{\pi d^2 C_m}{4}\sqrt{\frac{2}{\rho(1-\overline{\beta}^2)}} \tag{9}$$

$C_m$ is the nozzle circulation parameter, depending on its selection specifications (Jiang et al., 2021), $\overline{\beta}$ is the spool circulation area ratio under the steady-state operating point, ρ is the fluid density, with $d$ being the internal diameter of the pipeline. According to the above steady-state model and table 1 of layer characteristics parameters, the relationship curve between the pressure after the valve $P$ and the flow of each layer section or the flow of the whole well shown in Fig. 4.

**Table 1 Table of layer parameters**

| Item | Water absorption index ($k^{-1}$) $m^3 \cdot h^{-1} \cdot MPa^{-1}$ | Opening pressure ($b$) MPa | Opening ($\overline{\beta}$) % |
|---|---|---|---|
| **1** | 2.23 | 1.35 | 35 |
| **2** | 1.12 | 2.07 | 50 |
| **3** | 3.96 | 3.45 | 60 |

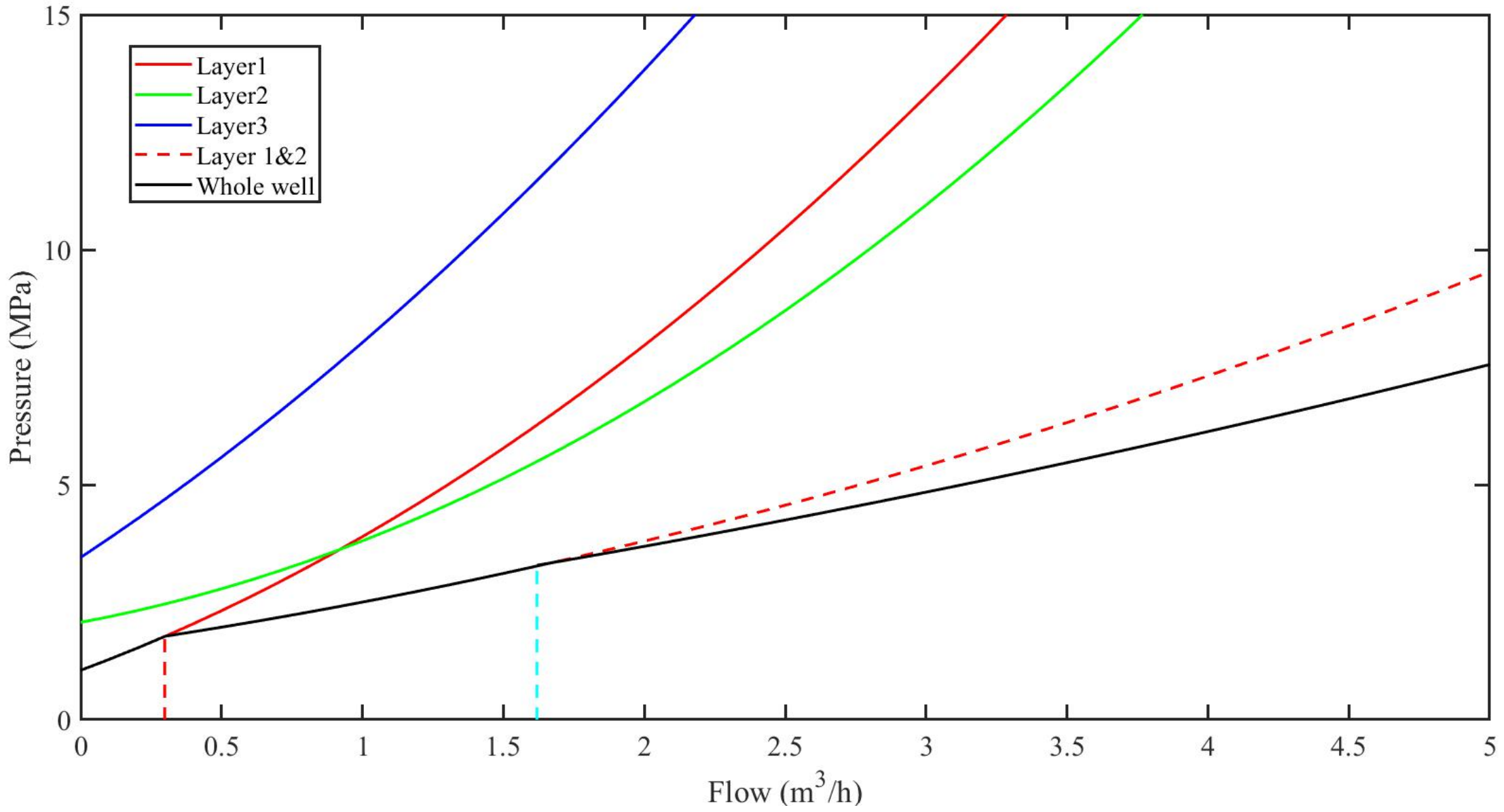

**Fig 4 Steady-state simulation results of pressure versus flow after the valve**

The curve in the figure is generated by connecting the points of the whole well. When the system is at the continuous operating point, the nozzle action is extremely slow and each dynamic process cycle is extremely short, so the pressure and flow value changes can be regarded as steady-state processes. The colored curves show the pressure-flow relationship for each layer segment, while the black curve at the bottom represents the full well flow versus pressure with openness during a steady state. From the figure, it can be seen that the curve of the high permeability layer is flatter, and its flow value varies more under the same change of opening degree, and the injection efficiency is higher. As the post-valve pressure increases, the high opening pressure layer section begins to meet the conditions for injection flow, so the full well flow-pressure curve shows a segmented trend, and the trend becomes smoother, while the wave code communication becomes more difficult. Therefore, in practice, in order to maintain the water injection qualification rate of low-permeability reservoirs while ensuring the effectiveness of wave code communication in each layer, it is necessary to close the water nozzles of certain high-permeability layers to balance the flow between layers. In wave code communication, the underground water distributor drives the underground nozzle opening degree to change periodically, generating wave code to complete the data upload operation. The fluctuation range of nozzle opening in the selected 2-layer section is 50~75%, corresponding to the pressure-flow relationship curve after the valve as shown in Fig. 5.

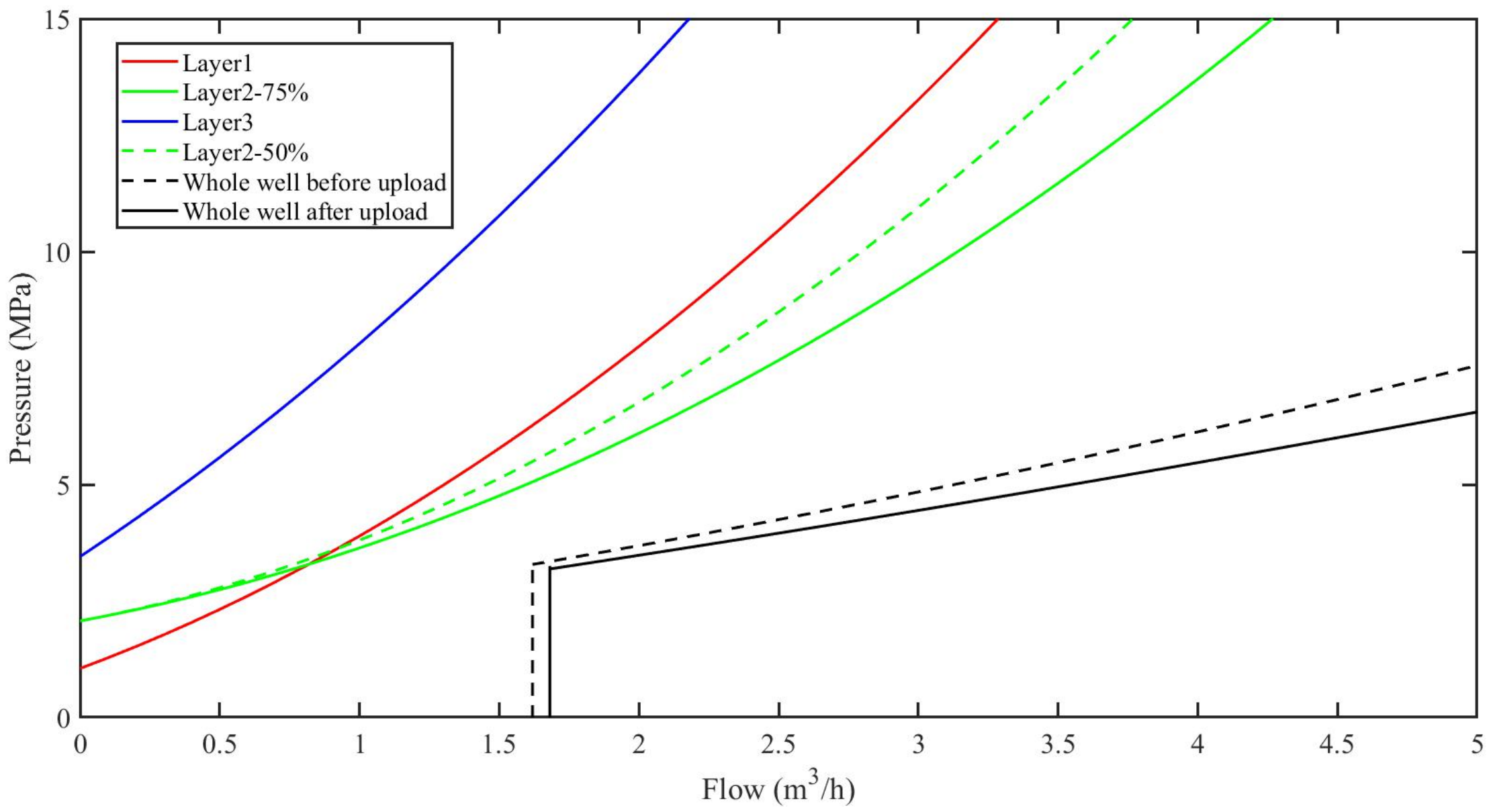

**Fig 5 Steady-state simulation results of pressure versus flow value after valve under data upload in high permeability section**

## 2.2 Dynamic model

The previous chapter analyzed the steady-state operating points of each layer segment and solved the wave code limit amplitude, but the description of the dynamic generation process of wave code is lacking, and the mechanism of wave code generation cannot be defined. Combining the actuator characteristics with the wave code transmission characteristics, it is necessary to model the fluid fluctuation dynamics under the fast control of the underground water distributor. In consideration of the working condition that the ground valve opening is fixed and the underground water distributor drives the nozzle set regulation, the dynamic flow response characteristics of the single-seat valve movement in the $i$-th layer section are described as following equation to define the dynamic process of regulation:

$$m\ddot{q}_{pe(i)}(t)+\mu\dot{q}_{pe(i)}(t)+\varepsilon q_{pe(i)}(t)=K_{v(i)}\beta_{(i)}(t)F_{m(i)}(t-\tau_c) \tag{10}$$

$q_{pe}$ is the relative flow and $K_v$ is the gain factor derived from the valve's algebraic flow characteristics of the valve.

$$K_v=R^{(\beta^{-1}-1)}lnR \tag{11}$$

$R$ is the adjustable ratio of nozzle flow, $m$,$\varepsilon$ is the mechanical parameters associated with the nozzle valve body, $\mu$ is the coefficient of friction of media viscosity, $\beta$ is the flow ratio of the nozzle, $F_m$ is the spool force when the controller signal is fully loaded. $\tau_c$ is the delay time of valve movement, which is related to the nozzle stem turning angle speed and signal delay time, and $t_s$ is the sampling period. To establish the relationship between the steady-state and dynamic models. The above relationship is combined to obtain the steady-state flow operating point of the layer whose nozzle acts, which leads to the full process characteristic model of this layer as following equation:

$$m\ddot{q}_{v(i)}(t)+\mu\dot{q}_{v(i)}(t)+\varepsilon q_{v(i)}(t)=K_{v(i)}q_{v(i)}\beta_{(i)}F_{m(i)}(t-\tau_c) \tag{12}$$

In addition, due to the different pipeline flow in a layer nozzle action will produce interlayer mutual disturbance effect, according to the principle of flow mutual disturbance process, the relative flow $q_{pe}$ is directly related to disturbance target layer's nozzle opening degree $\beta_{(j)}$, and the entire action process $\beta_{(j)}$ is constant, so the reciprocal layer section flow change is essentially achieved by changing the pressure after the valve $P$. The change of flow or pressure will be propagated to other layers after a certain period of time in the form of energy waves.

Set the wave code water strike propagation velocity for $v$, which is related to the pipe diameter, signal frequency and other parameters. Set $l_e$ as the unit tube length, the difference in well depth between neighboring layers is $h_d$, then the dynamic process model for the action of a single-layer section nozzle on the output of the jth layer section is described as follows:

$$[m\ddot{q}_{v(ij)}(t)+\mu\dot{q}_{v(ij)}(t)+\varepsilon q_{v(ij)}(t)][l_e\dot{q}_{v(ij)}(t)+vq_{v(ij)}(t)] \\ =vK_{v(j)}q_{v(ij)}\beta_{(j)}F_{m(j)}(t-\tau_c-\tau_l) \tag{13}$$

$\tau_l$ is the path delay for wave code transmission:

$$\tau_l=|i-j|\tau_e,\tau_e=\frac{h_d}{v} \tag{14}$$

Extending the above dynamic model into a state-space form as:

$$\dot{\boldsymbol{x}}_{\boldsymbol{m}}=\boldsymbol{diag}(\boldsymbol{A}_{\boldsymbol{me}})\boldsymbol{x}_{\boldsymbol{m}}+\boldsymbol{diag}(\boldsymbol{B}_{\boldsymbol{me}})\boldsymbol{u} \\ \boldsymbol{y}(k)=\boldsymbol{K}_{\boldsymbol{q}}*\boldsymbol{C}_{\boldsymbol{me}}\boldsymbol{x}_{\boldsymbol{m}} \tag{15a}$$

$$\boldsymbol{A}_{\boldsymbol{me}}=\begin{bmatrix} & \boldsymbol{0} & & & \boldsymbol{I}_{\boldsymbol{n-1}} & & \\ -\varepsilon & -\mu-\varepsilon\tau_e & -g_3 & \cdots & -g_{n+1} & mn\tau_e^{n-1}+\mu\tau_e^n & -m\tau_e^n \end{bmatrix},\boldsymbol{B}_{\boldsymbol{me}}=\begin{bmatrix}0\\ \vdots\\ 0\\ 1\end{bmatrix}_{1*n} \tag{15b}$$

$$\boldsymbol{C}_{\boldsymbol{me}}=\begin{bmatrix}\boldsymbol{C}_{\boldsymbol{me1}}\\ \vdots\\ \boldsymbol{C}_{\boldsymbol{meN}}\end{bmatrix}^{\boldsymbol{T}},\boldsymbol{C}_{\boldsymbol{mej}}=\begin{bmatrix} 1 & C_{n-j+1}^1\tau_e & C_{n-j+1}^2\tau_e^2 & \cdots & C_{n-j+1}^{n-j+1}\tau_e^{n-j+1} & 0 & \cdots & 0\\ 1 & C_{n-j+2}^1\tau_e & C_{n-j+2}^2\tau_e^2 & \cdots & C_{n-j+2}^{n-j+2}\tau_e^{n-j+2} & 0 & \cdots & 0\\ \vdots & \vdots & & & \vdots & & & \\ 1 & C_n^1\tau_e & C_n^2\tau_e^2 & \cdots & C_n^{n-2}\tau_e^{n-2} & C_n^{n-1}\tau_e^{n-1} & \tau_e^n & \\ 1 & \vdots & \vdots & & \vdots & \vdots & \ddots & 0\\ 1 & C_{n-j+N}^1\tau_e & C_{n-j+N}^2\tau_e^2 & \cdots & C_{n-j+N}^{n-j+N}\tau_e^{n-j+N} & 0 & \cdots & 0 \end{bmatrix}$$

$g$ is the coefficient of combination for the dynamic characteristics of the equipment:

$$g_j=C_n^{n-j+3}m\tau_e^{j-3}+C_n^{n-j+2}\tau_e^{j-2}+C_n^{n-j+1}\varepsilon\tau_e^{j-1},3\le j\le n+1,n=\frac{N(N-1)}{2},N\ge 2 \tag{15c}$$

$\boldsymbol{K}_{\boldsymbol{q}}$ is the steady-state operating point coefficient matrix:

$$\boldsymbol{K}_{\boldsymbol{q}}=\boldsymbol{K}_{\boldsymbol{v}}*\overline{\boldsymbol{q}_{\boldsymbol{v}}},\overline{\boldsymbol{q}_{\boldsymbol{v}}}=\begin{bmatrix}\overline{\boldsymbol{q}_{\boldsymbol{v}(1)}} & \cdots & \overline{\boldsymbol{q}_{\boldsymbol{v}(N1)}}\\ \vdots & \ddots & \vdots\\ \overline{\boldsymbol{q}_{\boldsymbol{v}(1N)}} & \cdots & \overline{\boldsymbol{q}_{\boldsymbol{v}(N)}}\end{bmatrix} \\ \overline{\boldsymbol{q}_{\boldsymbol{v}(ij)}}=\begin{bmatrix}\overline{q_{v(ij)}}\\ \vdots\\ \overline{q_{v(ij)}}\end{bmatrix}_{n*1}^{T},\boldsymbol{K}_{\boldsymbol{v}}=\begin{bmatrix}K_{v(1)} & \cdots & K_{v(1)}\\ \vdots & \ddots & \vdots\\ K_{v(N)} & \cdots & K_{v(N)}\end{bmatrix}_{N*(n*N)} \tag{15d}$$

The state-space matrices of the controlled dynamic model can be obtained as：

$$\boldsymbol{A}_{\boldsymbol{m}(n*N)*(n*N)}=\boldsymbol{diag}(\boldsymbol{A}_{\boldsymbol{me}}),\boldsymbol{B}_{\boldsymbol{m}(n*N)*1}=\boldsymbol{diag}(\boldsymbol{B}_{\boldsymbol{me}}),\boldsymbol{C}_{\boldsymbol{m}}=\boldsymbol{K}_{\boldsymbol{q}}*\boldsymbol{C}_{\boldsymbol{me}} \tag{15e}$$

The set of state vectors and the set of input and output vectors are shown below：

$$\boldsymbol{u}(k)=\begin{bmatrix}\beta_{(1)}\\ \beta_{(2)}\\ \vdots\\ \beta_{(N)}\end{bmatrix},\boldsymbol{x}_{\boldsymbol{m}}(k)=\begin{bmatrix}\boldsymbol{q}_{\boldsymbol{e}(\boldsymbol{1})}\\ \boldsymbol{q}_{\boldsymbol{e}(\boldsymbol{2})}\\ \vdots\\ \boldsymbol{q}_{\boldsymbol{e}(\boldsymbol{N})}\end{bmatrix},\boldsymbol{q}_{\boldsymbol{e}(\boldsymbol{i})}=\begin{bmatrix}q_{e(i1)}\\ q_{e(i2)}\\ \vdots\\ q_{e(ii)}\\ \vdots\\ q_{e(in)}\end{bmatrix},\boldsymbol{y}(k)=\begin{bmatrix}q_{v(1)}\\ \vdots\\ q_{v(N)}\end{bmatrix} \tag{15f}$$

Where $\boldsymbol{\beta}$ is the corresponding flow ratio of each layer segment nozzle, $q_e$ in $\boldsymbol{x}_{\boldsymbol{m}}$ represents the flow hysteresis change triggered by unit nozzle opening change, and the output variable $q_v$ represents the layer segment flow. The construction of the open-loop model of the layered water injection wave code generation process has been completed.

# 3 Fusion control algorithm

To optimize the dynamic response performance of layered water injection wave code control for efficient transmission communication, the MPC-PID-based weighted fusion control algorithm is introduced. The MPC controller is in the outer loop of the algorithm structure, which is responsible for providing reference values for prediction optimization, ensuring the traceability and accuracy of the system output, and releasing the coupling effect of single-loop parameter changes on other loops, thus realizing independent optimization of wave code control in each layer. The cascade PID control link is placed in the inner loop, which has a simple structure and fast operation speed, which weakens the disadvantage of slow MPC operation under the complex model to improve the wave code generation speed and suppress the sudden disturbance in the downhole transmission environment. Finally, the optimal fusion of the dual algorithm output law is performed to further enhance the quality of wave code generation and the reliability of wave code communication under abnormal operating conditions.

### 3.1 Multivariate model predictive control

This section introduces the principles of the multivariate model predictive control algorithm. The predictive controller consists of three internal components: a known iterative model, finite time-domain optimization, and error feedback correction. A centralized parameter approach is used to control the multivariate model, and the multi-loop modified model response matrix is first loaded into the controller in the following form:

$$\boldsymbol{A}=\begin{bmatrix}\boldsymbol{A_{11}} & \cdots & \boldsymbol{A_{1,mimo_i}}\\ \vdots & \ddots & \vdots\\ \boldsymbol{A_{mimo_o,1}} & \cdots & \boldsymbol{A_{mimo_o,mimo_i}}\end{bmatrix},\boldsymbol{A_{ij}}=\begin{bmatrix}a_{ij(1)} & 0 \quad \cdots & 0\\ a_{ij(2)} & a_{ij(1)} & 0\\ \vdots & \ddots & \vdots\\ a_{ij(p)} & \cdots & a_{ij(p-m+1)}\end{bmatrix} \tag{16a}$$

$a_{ij(n)}$ is the $n$th step discrete response of the $j$th input corresponding to the $i$th output. $p$ and $m$ are the prediction step and control step respectively. Input multivariate reference trajectory matrix $\boldsymbol{R_w(k)}$ as the setpoint for each loop and the real-time control output is optimal using the minimization cost function as follow:

$$\Delta\boldsymbol{u}^*(k)=\boldsymbol{L^T}(\boldsymbol{A^T Q A}+\boldsymbol{R})^{-1}\boldsymbol{A}^T\boldsymbol{Q}[\boldsymbol{R_w}(k)-\boldsymbol{y_{\tilde{N}0}}(k)] \tag{16b}$$

where $\boldsymbol{L^T}$ is obtained by expanding the unit array, representing the first term of the optimal control output matrix taken. $\boldsymbol{Q}$ is the error weight matrix and $\boldsymbol{R}$ is the control weight matrix, the error vector values for each step of the same loop are selected to have the same factors of the following form:

$$\boldsymbol{Q}=\begin{bmatrix}\begin{matrix}Q_1 & \dots & 0\\ \vdots & \ddots & \vdots\\ 0 & \dots & Q_1\end{matrix} & \cdots & 0\\ \vdots & \ddots & \vdots\\ 0 & \cdots & \begin{matrix}Q_{mimo_o} & \cdots & 0\\ \vdots & \ddots & \vdots\\ 0 & \cdots & Q_{mimo_o}\end{matrix}\end{bmatrix},\boldsymbol{R}=\begin{bmatrix}\begin{matrix}R_1 & \dots & 0\\ \vdots & \ddots & \vdots\\ 0 & \dots & R_1\end{matrix} & \cdots & 0\\ \vdots & \ddots & \vdots\\ 0 & \cdots & \begin{matrix}R_{mimo_i} & \cdots & 0\\ \vdots & \ddots & \vdots\\ 0 & \cdots & R_{mimo_i}\end{matrix}\end{bmatrix} \tag{16c}$$

$mimo_i$ represents the number of input variables and $mimo_o$ represents the number of output variables. From the model structure, it is obtained that:

$$mimo_i=mimo_o=N \tag{17}$$

As the system is coupled, the control output is equal to a linear superposition of the inputs. The predicted output is obtained by computing the control output with the model response matrix $\boldsymbol{a_N}$, and the error vector $\boldsymbol{e}(k+1)$ is divided by the actual value and feeding it back to obtain the corrected output. Setting the covariance coefficients of the non-first step in the feedback matrix of the same loop uniformly.

$$\boldsymbol{H}=\begin{bmatrix}\boldsymbol{h_1} & \cdots & 0\\ \vdots & \ddots & \vdots\\ 0 & \cdots & \boldsymbol{h_{mimo_o}}\end{bmatrix},\boldsymbol{h_t}=\begin{bmatrix}1\\ h_t\\ h_t\\ \vdots\\ h_t\end{bmatrix} \tag{18}$$

The correction values are shifted to obtain the first step prediction matrix for $k+1$ steps of the following form:

$$\boldsymbol{y}_{\widetilde{N0}}(k+1) = \boldsymbol{S}[\boldsymbol{y}_{\widetilde{N0}}(k) + \boldsymbol{a}_N \Delta \boldsymbol{u}^*(k) + \boldsymbol{He}(k+1)] \tag{19}$$

The displacement matrix $\boldsymbol{S}$ is a diagonal array formed by $mimo_o$ shift matrices, and the value of each shift matrix element satisfies $\boldsymbol{S}_{i,i+1} = \boldsymbol{S}_{N,N} = 1$. According to the above principle, the error weight matrix $\boldsymbol{Q}$ is in charge of strengthening the output of the system, while the control weight matrix $\boldsymbol{R}$ inhibits response output (Eugenio et al., 2019). The feedback vector matrix is directly related to the dynamic characteristics and stability of the system, and the feedback value is proportional to the response speed of the system within a certain limit, which can lead to system dispersion in case of excess. In the computation of the optimal control law, the coupling terms of the model are eliminated by algebraic computation, yet the actual coupling effect exists. However, the parameter tuning of one loop in the controller section does not affect the output of other loops, so the multivariate model predictive control has a certain internal decoupling capability.

## 3.2 Output fusion control method

The concrete fusion control algorithm is described in this chapter. The discrete PID controller is placed in the inner loop of the system, and the MPC controller provides the master setpoint. Since the PID algorithm is less effective in controlling the coupled system, a decoupler needs to be added to the inner loop. The multivariate coupled model is processed using a dynamic decoupling algorithm, which gives the input transformation array $\boldsymbol{M}$ and the state feedback matrix $\boldsymbol{K}$ of the decoupled system based on the state-space model.

$$\boldsymbol{K} = \boldsymbol{M}\widetilde{\boldsymbol{K}}\boldsymbol{T} + \boldsymbol{MF}, \boldsymbol{M} = \begin{bmatrix} \boldsymbol{C}_{m1j}\boldsymbol{A}_m\boldsymbol{B}_m \\ \boldsymbol{C}_{m2j}\boldsymbol{A}_m\boldsymbol{B}_m \\ \vdots \\ \boldsymbol{C}_{mNj}\boldsymbol{A}_m\boldsymbol{B}_m \end{bmatrix}^{-1}, \boldsymbol{F} = \begin{bmatrix} \boldsymbol{C}_{m1j}\boldsymbol{A}_m^{mimo_o} \\ \boldsymbol{C}_{m2j}\boldsymbol{A}_m^{mimo_o} \\ \vdots \\ \boldsymbol{C}_{mNj}\boldsymbol{A}_m^{mimo_o} \end{bmatrix}^{-1} \tag{20a}$$

$\boldsymbol{T}$ in the above equation is the decoupled canonical transformation matrix, which satisfies the conditions as follow:

$$\boldsymbol{T}^{-1}(\boldsymbol{A} - \boldsymbol{BMF})\boldsymbol{T} = \boldsymbol{A}^*,\ \boldsymbol{T}^{-1}\boldsymbol{BM} = \boldsymbol{B}^*,\ \boldsymbol{CT} = \boldsymbol{C}^* \tag{20b}$$

$\widetilde{\boldsymbol{K}}$ is the pole configuration matrix, which is typically set in the negative half-plane to ensure the stability of the system. The state-space model matrix after the coupling action disappears is shown as follows:

$$\boldsymbol{A}_{\wedge} = \boldsymbol{A}_m - \boldsymbol{B}_m\boldsymbol{K}, \boldsymbol{B}_{\wedge} = \boldsymbol{MB}_m, \boldsymbol{C}_{\wedge} = \boldsymbol{C}_m \tag{20c}$$

Set the cascade partial weight coefficient array to be $\boldsymbol{w}$. Since MPC control does not require decoupling, its output is applied directly to the object and given the weight coefficient $\boldsymbol{w}'$. The algebraic relations exist between the weight matrices can be obtained as.

$$\boldsymbol{w} + \boldsymbol{w}' = \boldsymbol{I}_{mimo_o*1} \tag{21}$$

Therefore, the fusion controller consists of a PID controller, an assignment session, and a centralized MPC controller, the control output can be expressed as:

$$\boldsymbol{u} = \boldsymbol{w}\boldsymbol{u}_c + \boldsymbol{w}'\boldsymbol{u}_m = \boldsymbol{w}_1\boldsymbol{u}_c + (\boldsymbol{I}_{mimo_o*1} - \boldsymbol{w})\boldsymbol{u}_m \tag{22}$$

Since the output of the PID control algorithm is not dependent on the model, it can be equated to an independent model with multiple parallel connections and as part of the generalized controlled system. The associated state variables of the PID controller matrix are denoted as $[A_p, B_p, ...]$. The target output of the generalized system can be obtained from the linear operation property of the state-space system as follows:

$$\dot{\boldsymbol{x}}_0 = \boldsymbol{A}\boldsymbol{x}_0 + \boldsymbol{B}\boldsymbol{u}, \boldsymbol{y}_0 = \boldsymbol{C}\boldsymbol{x}_0 + \boldsymbol{D}\boldsymbol{u} \tag{23a}$$

The generalized system is divided into two parts by the basic assignment calculation, and the corresponding weights of each part correspond to the weights of the control algorithm. The topological model of the multivariate fusion controller and the structure of the generalized system model are shown in Fig. 6 and Fig. 7 respectively.

According to the above analysis, the state-space model of the generalized system under MPC control can be derived as follows:

$$\boldsymbol{A}=\begin{bmatrix}\boldsymbol{A_p} & -\boldsymbol{B_p C_m w} & \boldsymbol{0}\\ \boldsymbol{B_m M C_p} & \boldsymbol{A_m}-\boldsymbol{B_m K}-\boldsymbol{B_m M D_p C_m w} & \boldsymbol{0}\\ \boldsymbol{0} & -\boldsymbol{B_m C_m w} & \boldsymbol{A_m}\end{bmatrix},\boldsymbol{B}=\begin{bmatrix}\boldsymbol{B_p}\\ \boldsymbol{B_m M D_p}\\ \boldsymbol{B_m}\end{bmatrix} \tag{23b}$$

$$\boldsymbol{C}=[\boldsymbol{0}\quad \boldsymbol{C_m w}\quad \boldsymbol{C_m}(\boldsymbol{I}_{\boldsymbol{mimo_o*1}}-\boldsymbol{w})],\boldsymbol{D}=\boldsymbol{0}$$

where the set of state and control vectors are shown as follows:

$$\boldsymbol{x_0}=\begin{bmatrix}\boldsymbol{x_{po}}\\ \boldsymbol{x_{mo}}\\ \boldsymbol{x_{co}}\end{bmatrix},\boldsymbol{x_l}=\begin{bmatrix}x_{l(1)}\\ \cdots\\ x_{l(n*N)}\end{bmatrix},\{l=po,mo,\ldots\},\boldsymbol{u}=\begin{bmatrix}\beta_{(1)}\\ \cdots\\ \beta_{(N)}\end{bmatrix},\boldsymbol{y}=\begin{bmatrix}q_{v(1)}\\ \cdots\\ q_{v(N)}\end{bmatrix} \tag{23c}$$

As shown by the generalized system structure analysis, $\boldsymbol{x_{po}}$、$\boldsymbol{x_{mo}}$、$\boldsymbol{x_{co}}$ correspond to the outputs $\boldsymbol{y_{po}}$、$\boldsymbol{y_{mo}}$、$\boldsymbol{y_{co}}$ of the PID controller, the cascaded weighting part of the system, and the MPC weighting part of the system separately.

$$\boldsymbol{y_{po}}=\boldsymbol{C_p x_{po}}+\boldsymbol{D_p u},\boldsymbol{y_{mo}}=\boldsymbol{C_m x_{mo}},\boldsymbol{y_{co}}=\boldsymbol{C_m x_{co}} \tag{23d}$$

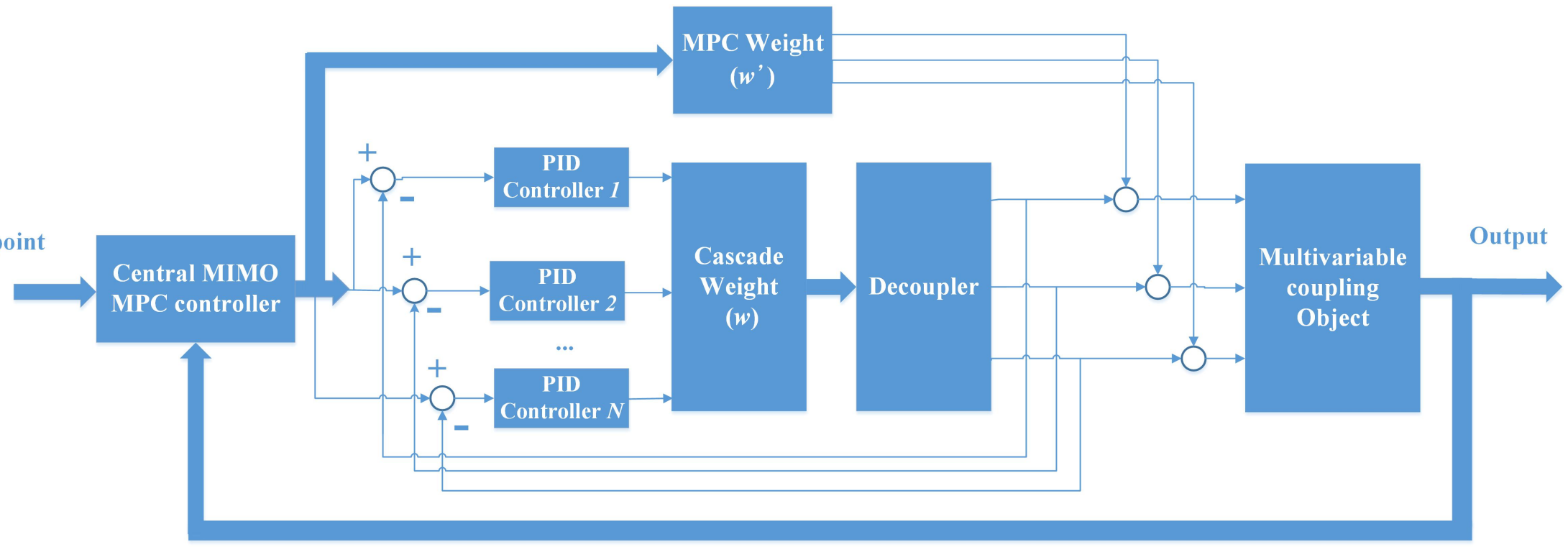

**Fig 6 Multivariable MPC-PID fusion controller structure**

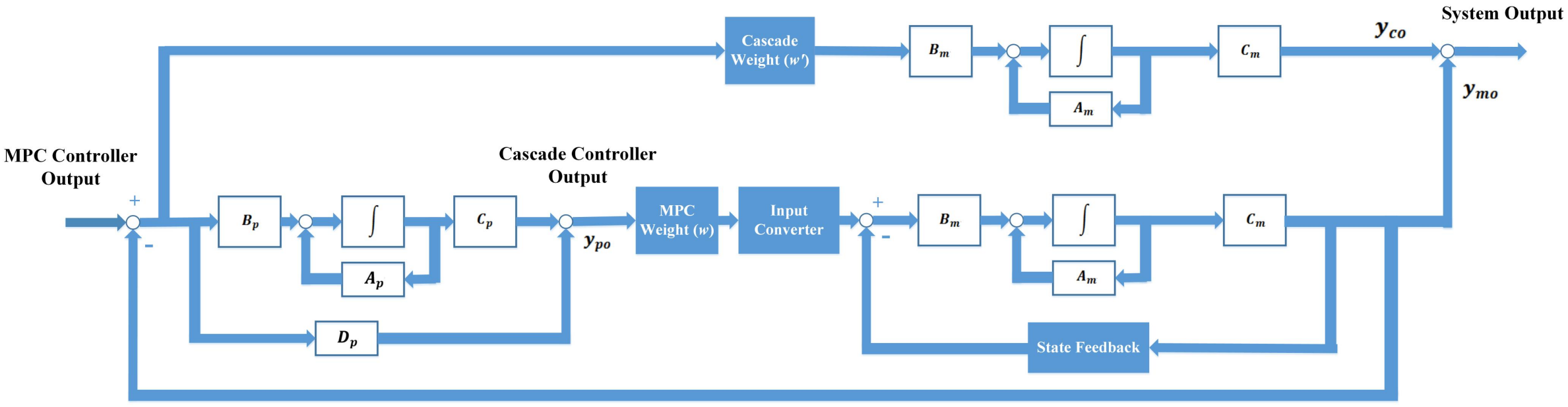

**Fig 7 Internal structure of the generalized system state-space model**

Then derive the optimal fusion output controller based on the above state-space model. Firstly, the system is discretized with a sampling period of $t_e$, which can be obtained:

$$\begin{aligned}\boldsymbol{x_0}(k+1)&=\boldsymbol{A_0 x_0}(k)+\boldsymbol{B_0 u}(k),\\ \boldsymbol{y_0}(k)&=\boldsymbol{C_0 x_0}(k)+\boldsymbol{D_0 u}(k)\end{aligned} \tag{24a}$$

The discrete system state vector after predicting $N_p$ steps at step 0 of the controller is as follow:

$$\widetilde{\boldsymbol{x_1}}(N_p|0)=\boldsymbol{\mathcal{A}_0 x}(0)+\boldsymbol{\varphi}\Delta\boldsymbol{u}(0) \tag{24b}$$

where $\boldsymbol{\mathcal{A}_0}$, $\boldsymbol{\varphi}$ are iterative state transfer vector arrays of the following form:

$$\mathcal{A}_0 = \begin{bmatrix} A_0 \\ A_0^2 \\ \vdots \\ A_0^{N_p} \end{bmatrix}, \varphi = \begin{bmatrix} B_0 & 0 & & 0 \\ A_0 B_0 & B_0 & \cdots & 0 \\ A_0^2 B_0 & A_0 B_0 & & 0 \\ \vdots & & \ddots & \vdots \\ A_0^{N_p-1} B_0 & A_0^{N_p-2} B_0 & \cdots & A_0^{N_p-N_m} B_0 \end{bmatrix} \tag{24c}$$

The predicted output of the zero-step according to the state-space model can be obtained as:

$$\begin{aligned} \widetilde{y_1}(N_p|0) &= \widetilde{y_0}(N_p|0) + C_0 \widetilde{x_1}(N_p|0) + D_0 \Delta u(0) \\ &= \widetilde{y_0}(N_p|0) + C_0 \mathcal{A}_0 x(0) + (C_0 \varphi + D_0) \Delta u(0), \widetilde{y_0}(N_p|0) = 0 \end{aligned} \tag{25}$$

Take the error cost function matrix of the following form：

$$J = \left\| \begin{bmatrix} [R_{w(1)}(N_p) - \widetilde{y_{1(1)}}(N_p|0)]^T \\ [R_{w(2)}(N_p) - \widetilde{y_{1(2)}}(N_p|0)]^T \\ \vdots \\ [R_{w(N)}(N_p) - \widetilde{y_{1(N)}}(N_p|0)]^T \end{bmatrix} Q \begin{bmatrix} R_{w(1)}(N_p) - \widetilde{y_{1(1)}}(N_p|0) \\ R_{w(2)}(N_p) - \widetilde{y_{1(2)}}(N_p|0) \\ \vdots \\ R_{w(N)}(N_p) - \widetilde{y_{1(N)}}(N_p|0) \end{bmatrix} \right\| + |\Delta u_M^T R \Delta u_M| \tag{26}$$

After minimizing the cost function, the optimal predictive control increment can be obtained as：

$$\Delta u_M = L^T[(C_0\varphi + D_0)^T Q (C_0\varphi + D_0) + R]^{-1} (C_0\varphi + D_0)^T Q [R_w(N_p) - C_0 \mathcal{A}_0 x(0)] \tag{27a}$$

Further, the discrete control law for the fusion controller with fixed weights is solved as:

$$u^*(z) = [(1 - z^{-1})^{-1} \Delta u_M] \left\{ I + w \left[ C_p (zI - e^{A_p t_e})^{-1} \sum_{k=0}^{\infty} [(k+1)!]^{-1} A_p^k t_e^{k+1} B_p + D_p - I \right] \right\} \tag{27b}$$

The real output $y_1$ obtained from the control law input yields the following vector matrix of first-step predicted values for the next step.

$$\widetilde{y_0}(N_p + 1|1) = S[\widetilde{y_1}(N_p|0) + H(y_1(N_p|0) - \widetilde{y_1}(N_p|0)] \tag{28}$$

Replacing the first predicted value obtained by the loop into equation (25) completes the cyclic prediction output of the fusion control.

To optimize the dynamic response performance of layered water injection wave code control for efficient transmission communication, the MPC-PID-based weighted fusion control algorithm is introduced. The MPC controller is in the outer loop of the algorithm structure, which is responsible for providing reference values for prediction optimization, ensuring the traceability and accuracy of the system output, and releasing the coupling effect of single-loop parameter changes on other loops, thus realizing independent optimization of wave code control in each layer. The cascade PID control link is placed in the inner loop, which has a simple structure and fast operation speed, which weakens the disadvantage of slow MPC operation under the complex model to improve the wave code generation speed and suppress the sudden disturbance in the downhole transmission environment. Finally, the optimal fusion of the dual algorithm output law is performed to further enhance the quality of wave code generation and the reliability of wave code communication under abnormal operating conditions.

## 3.3 Optimal configuration of weights

The theoretical derivation of the algorithmic principle of output fusion control in the previous section shows that the controller structure is more complex, and the role of weights on the overall closed-loop system is uncertain. To deeply study the fusion mechanism of MPC and PID control algorithms and dissect the performance state of the wave code dynamic response under the change of weights, it is necessary to study the analytical method of the optimal weights of the fusion control. The analysis of the internal model structure of the fusion control is first performed. The transformation of the internal model structure aims to describe the system characteristics using a generalized frequency domain transfer function, which in turn visualizes the quantitative relationship between the weights and the output.

To obtain the internal mode structure parameters, the model structure after fusion needs to be specified. According to the first prediction link of the fusion controller, it can be derived that:

$$\widetilde{\boldsymbol{y_0}}(N_p+1|1)=z\widetilde{\boldsymbol{y_1}}(N_p|0)=\boldsymbol{S}[\widetilde{\boldsymbol{y_1}}(N_p|0)+\boldsymbol{\varphi \Delta u_M}(z)+z\boldsymbol{He}(z)] \tag{29}$$

A feedback error structure of the following form can also be derived as:

$$\boldsymbol{e}(z)=\boldsymbol{y}(z)-[\boldsymbol{L^T\varphi}+\boldsymbol{L^T}(z\boldsymbol{I}-\boldsymbol{S})^{-1}\boldsymbol{\varphi}]\Delta \boldsymbol{u_M}(z)-z\boldsymbol{L^T}(z\boldsymbol{I}-\boldsymbol{S})^{-1}\boldsymbol{SHe}(z) \tag{30}$$

The equivalent controller $\boldsymbol{G_c}$ of the internal mode structure, the internal model $\boldsymbol{G}^{\wedge}$ and the equivalent feedback filter $\boldsymbol{G_f}$ can be obtained as:

$$\begin{gathered}\boldsymbol{G_c}(z)=[1+(z-1)^{-1}][\boldsymbol{I}+\boldsymbol{I}_{N*N}\boldsymbol{\phi}(z\boldsymbol{I}-\boldsymbol{S})^{-1}\boldsymbol{S\varphi}]^{-1}\boldsymbol{\phi}\\ \boldsymbol{G}^{\wedge}(z)=z^{-2}(z-1)\boldsymbol{L^T}[\boldsymbol{I}+(z\boldsymbol{I}-\boldsymbol{S})^{-1}\boldsymbol{S}]\boldsymbol{\varphi}\\ \boldsymbol{G_f}(z)=z[\boldsymbol{I}+\boldsymbol{L^T}(z\boldsymbol{I}-\boldsymbol{S})^{-1}\boldsymbol{SH}]^{-1}\boldsymbol{I}_{N*N}\boldsymbol{\phi}(z\boldsymbol{I}-\boldsymbol{S})^{-1}\boldsymbol{SH}\end{gathered} \tag{31a}$$

Where $\boldsymbol{\phi}$ is the optimization coefficient matrix:

$$\boldsymbol{\phi}=\boldsymbol{L^T}[(\boldsymbol{C_0\varphi}+\boldsymbol{D_0})^{\boldsymbol{T}}\boldsymbol{Q}(\boldsymbol{C_0\varphi}+\boldsymbol{D_0})+\boldsymbol{R}]^{-1}(\boldsymbol{C_0\varphi}+\boldsymbol{D_0})^{\boldsymbol{T}}\boldsymbol{Q} \tag{31b}$$

The structure of the fusion control internal mode can be obtained as shown in Fig. 8.

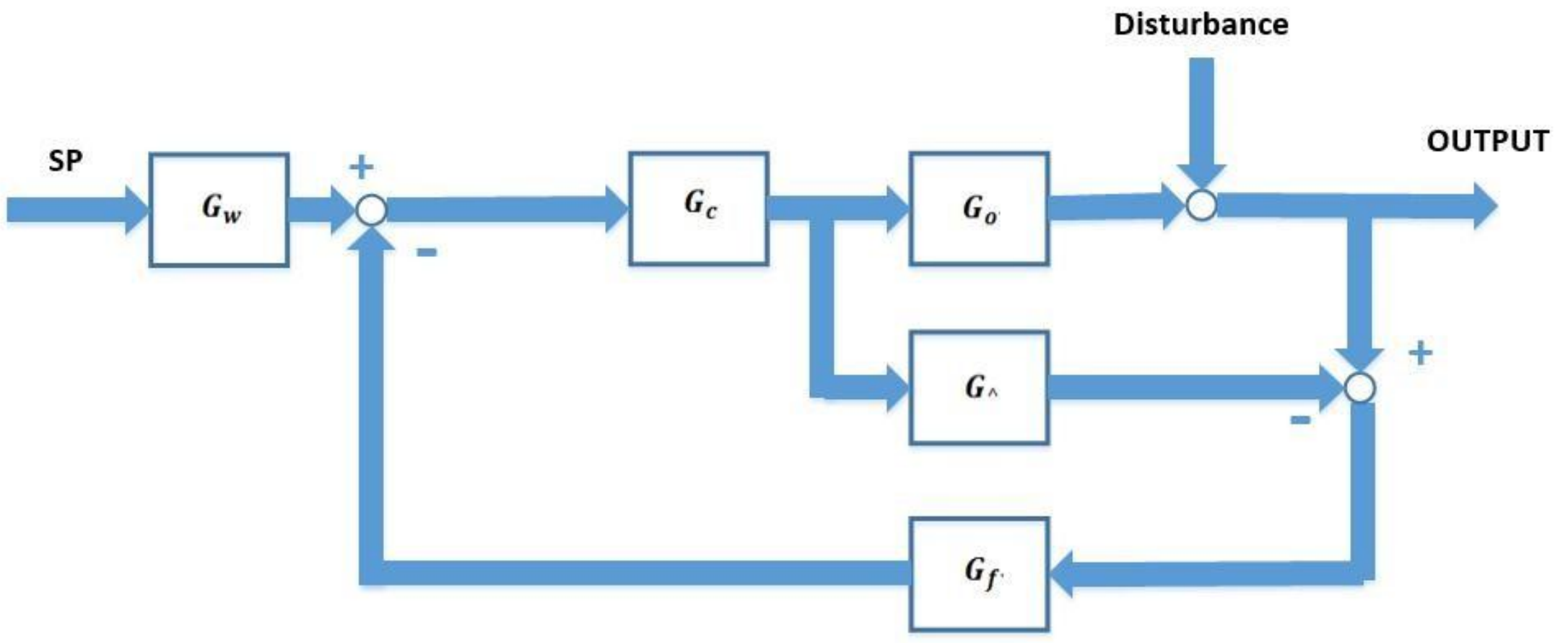

**Fig 8 Fusion control of the internal model structure**

This leads to the total closed-loop gain of the fusion control under the internal mode structure as:

$$\boldsymbol{K}(z)=\frac{\boldsymbol{G_0}(z)\boldsymbol{G_c}(z)}{\boldsymbol{I}+\boldsymbol{G_c}(z)\boldsymbol{G_f}(z)[\boldsymbol{G_0}(z)-\boldsymbol{G}^{\wedge}(z)]} \tag{32}$$

After using the final value theorem, the steady-state error expression can be derived as:

$$\boldsymbol{e_{ss}}=lim_{z\to 1}(z-1)\boldsymbol{K}=\boldsymbol{0} \tag{33}$$

Therefore, for a multivariable fusion control system with stable output, there is no steady-state residual error in the output of each loop. The above analysis shows that the total closed-loop gain model is of a high order, and to reduce the computational effort in engineering applications, it is necessary to reduce the order. To make the system realize the condition of degradable processing and reduce the approximation error, the dominant poles of the two models should be made similar, so the differential and integral time parameters of the controller can be roughly adjusted to keep the output oscillation of the original model within a rational range. When the system oscillation is too violent, the control effect will be seriously deteriorated, resulting in low quality of the control generated wave code and decoding failure. The specific approximate effect of the output performance of the continuous and discrete systems with a sampling period of 1s is shown in Fig. 9.

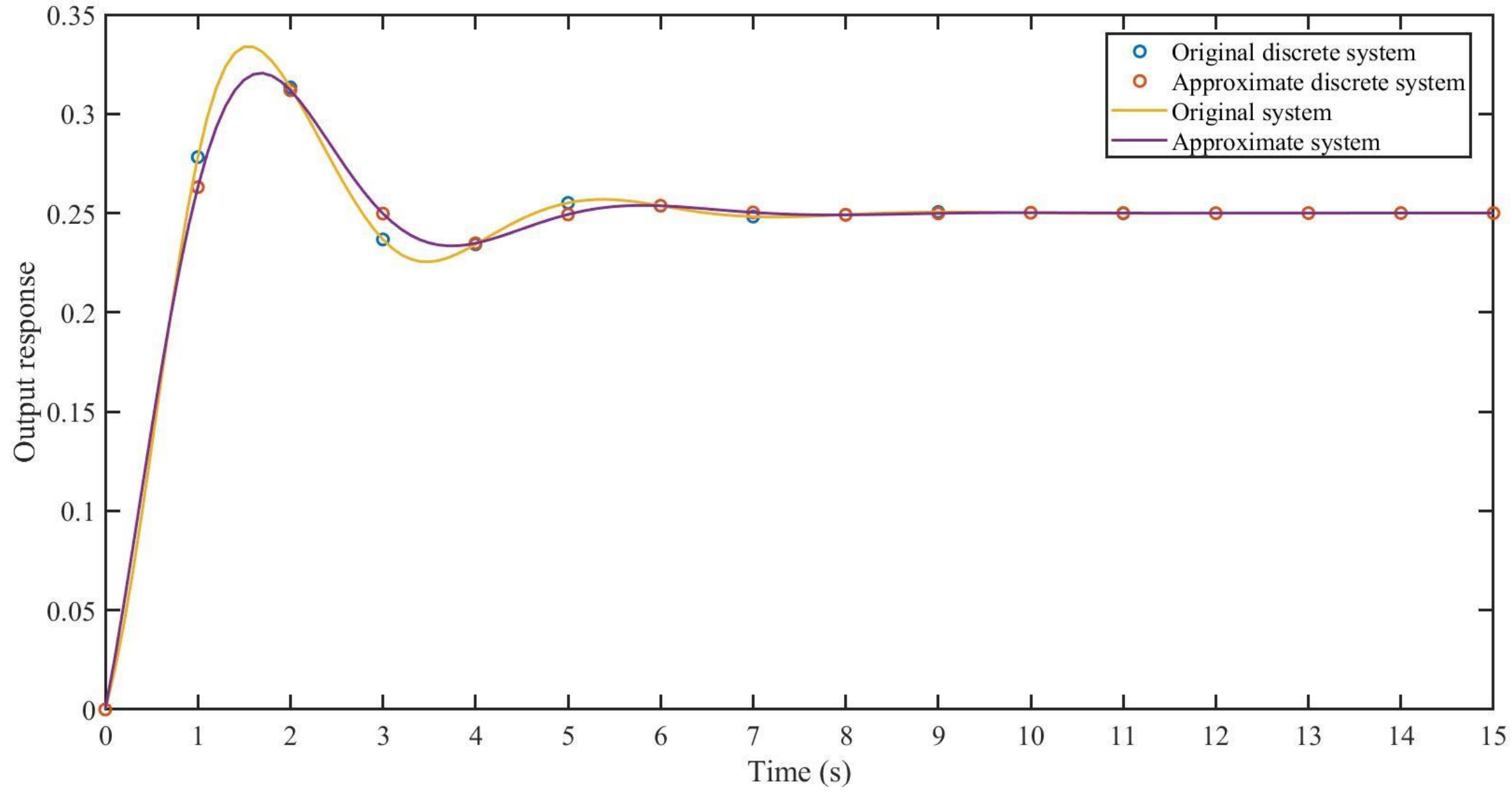

**Fig 9 Equivalent reduced-order approximation-time domain response effect**

By analyzing the frequency domain model of the system derived in the previous section and combining time-domain response of the fusion control, it is known that the output curve under the optimal weights is at the critical point where the first peak disappears, according to which the overall scheme of the optimal weights can be summarized as follows: using the frequency domain model of internal modulation to find the time domain expression of the output curve, this formula should be a function $f(t,w)$ about the weights $w$ and time $t$, and the output all peak moments are expressed as a function of the weights , so the first peak moment $t_p$ is solved as follows:

$$T=\left\{\frac{d(f(t,w))}{dt}=0|w\right\},\frac{d(f(t,w^*))}{dt}|_{t=t_p}=0,t_p=min\,T \qquad (34a)$$

Since the vibration curve disappears under the optimal weight and the output reaches the set value directly after the rise, the first peak value is in the critical disappearance state, and the peak value is equal to the steady-state output value at the moment of $t_s$ at this time, according to which it is obtained that:

$$f(t_p)-f[(t_p)+N_{ps}t_e]=0,N_{ps}\in(\frac{t_p}{t_e},\frac{t_s}{t_e}] \qquad (34b)$$

where $N_{ps}$ indicates the number of steps from peak to steady-state. Based on the above theory, the following algorithm for solving the optimal weight theory solution w^* can be derived by combining the simulation curves.

1.Performing a Tustin bilinear transformation on the controller part of the internal modalized output gain $\boldsymbol{K}$, where the $\boldsymbol{G_0}$ model part can be directly replaced using the following equation:

$$\boldsymbol{G_0}(s)=\boldsymbol{C_0}(s\boldsymbol{I}-\boldsymbol{A_0})^{-1}\boldsymbol{B_0}+\boldsymbol{D_0} \qquad (35)$$

$\boldsymbol{G_0}$ is the frequency domain model matrix of $mimo_o$ rows and $mimo_i$ columns, and the total output gain of $mimo_o$ terms is normalized to the following form:

$$\boldsymbol{K}(s)=\begin{bmatrix}K_1(s)\\ \vdots\\ K_{mimo_o}(s)\end{bmatrix}=\begin{bmatrix}[s^p+c_{1(1)}s^{p-1}+\ldots+c_{p(1)}]^{-1}[d_{1(1)}s^{p-1}+\ldots+d_{p(1)}]\\ \vdots\\ [s^p+c_{1(mimo_o)}s^{p-1}+\ldots+c_{p(mimo_o)}]^{-1}[d_{1(mimo_o)}s^{p-1}+\ldots+d_{p(mimo_o)}]\end{bmatrix} \qquad (36)$$

2. Using the equivalent transformation to downscale () and set the downscaled order as *r*. The output gain of the *t*-th term can be approximated as

$$\widetilde{\boldsymbol{K}}_{\boldsymbol{o}}(s)=\sum_{j=1}^{r}\beta_j\prod_{i(t)=2}^{j}[(\alpha_{i(t)}s+(\alpha_{i(t-1)}s)^{-1})^{-1}]^{(r)},r<\frac{p}{2} \tag{37}$$

where $\alpha,\beta$ are the weight correlation coefficients, which can be obtained by an iterative procedure, generally taking the calculated results after 2~3 iterations.

3.The frequency domain analysis of the reduced-order model is performed to obtain the weight-dependent output characteristics. To analyze the role played by the weights under each loop, it is necessary to categorize the discussion concerning the output characteristics: Fusion optimization should take reducing the overshoot as the primary target, and if the original string-level control output overshoot is small, the weight $w$ should be increased appropriately to retain the good dynamic performance of the lower-level PID. Conversely, if the original cascade control output oscillation is strong or the overshoot is large, the weight $w$ should be reduced to increase the robust performance of MPC control. Under the optimal weight, the first time the overshoot-free output curve reaches the peak is the steady-state, so the difference between the steady-state time and the first peak time must be extremely small, according to which the following weight solution equation can be derived as:

$$\begin{cases}\begin{cases}w_\sigma^*=max\,\mathcal{W}^*,\sigma(w)>0\\ \overline{w_\sigma^*}=min\,\mathcal{W}^*,\sigma(w)=0\end{cases}\\ \mathcal{W}^*=\begin{cases}argmin\left|\pi-\dfrac{2\beta_1(w)arctg\delta(w)}{2\beta_2(w)-\alpha_1(w)\beta_1(w)}-\dfrac{[8+2ln\sigma(w)]\delta(w)}{\alpha_1(w)}\right|,\sigma(w)>0\\ argmin\left|\pi-\dfrac{2\beta_1(w)arctg\delta(w)}{2\beta_2(w)-\alpha_1(w)\beta_1(w)}-\dfrac{8\delta(w)}{\alpha_1(w)}\right|,\sigma(w)=0\end{cases}\end{cases} \tag{38a}$$

Where $\delta$ is related to the output damping ratio of the fusion control concerning the vibration frequency, note that the system output is the first peak at the maximum overshoot, so $\delta$ takes the value in the first cycle.

$$\delta(w)=\sqrt{\alpha_2(w)-\frac{\alpha_1^2(w)}{4}},\delta(w)\in\left(-\frac{\pi}{2},\frac{\pi}{2}\right) \tag{38b}$$

$\sigma$ is the overshoot discriminant function.

$$\sigma(w)=\sqrt{1-\alpha_1(w)\frac{\beta_1(w)}{\beta_2(w)}+\alpha_2(w)[\frac{\beta_1(w)}{\beta_2(w)}]^2}|_{w=1} \tag{38c}$$

$\mathcal{W}^*$ represents the set of weight solutions without overshooting the output at the minimum difference between $t_p$ and $t_s$, so satisfying the $\mathcal{W}^*$ set basis constraint is a necessary and non-sufficient condition for weight optimization. Therefore, in order to balance the optimal performance of the dual algorithm, an analysis of the optimal selection of the solution set is required: if there is overshoot in the string-level output, * allows a small range of changes in the string-level weights $w_\sigma^*$ at this time, and at this time, in order to speed up the system response, the maximum value under the $\mathcal{W}^*$ constraint should be selected. On the other hand, if the original output does not have overshoot, the value range of $\overline{w_\sigma^*}$ is expanded. In order to compensate for the weakened MPC robust performance, the minimum value in the set should be taken. Therefore, under the same control loop, the original cascade output has overshoot. The optimal fusion weight $w_\sigma^*$ and the weight when the cascade output has no overshoot $\overline{w_\sigma^*}$ have the following relationship:

$$\mathcal{W}^*=(0,w_\sigma^*]\cup[\overline{w_\sigma^*},1),\overline{w_\sigma^*}>w_\sigma^* \tag{39}$$

In conclusion, the optimal system first needs to satisfy the primary performance constraint of overshoot-free output, while considering the secondary performance of response speed or robustness.

# 4 Simulation results analysis and discussion

## 4.1 Control Simulation and verification

In this section, a simulation scenario of a three-variable equivalent model for layered water injection is designed to show the response characteristics in the form of intuitive curves for different control algorithms for flow, pressure and related parameters of water distribution equipment under various operating conditions. Several evaluation metrics are also selected to objectively describe and compare the performance of different algorithms. The simulation compares the PID control algorithm, MPC control algorithm, MPC-PID cascade control algorithm, and fusion control algorithm with the target. The simulation operations were performed on a machine with Intel i5 CPU with 16G running memory. Firstly, the layer segment flow control simulation is conducted, in order to simulate various sudden actual working conditions, the sudden change link of the set value is introduced, and the step jump is set to be carried out asynchronous order to check the mutual disturbance effect of multi-loop control output change. The input setting signal curve of the control system is shown in Fig. 10. The flow value of the second layer section starts at $2m^3/h$ and then drops to $1m^3/h$ at 150s; the flow value of the second layer section is initially set at $1.5m^3/h$ and rises to $2.5m^3/h$ at 300s; the flow value of the third layer section is initially set at $3m^3/h$ and rises to $3.7m^3/h$ at 300s.

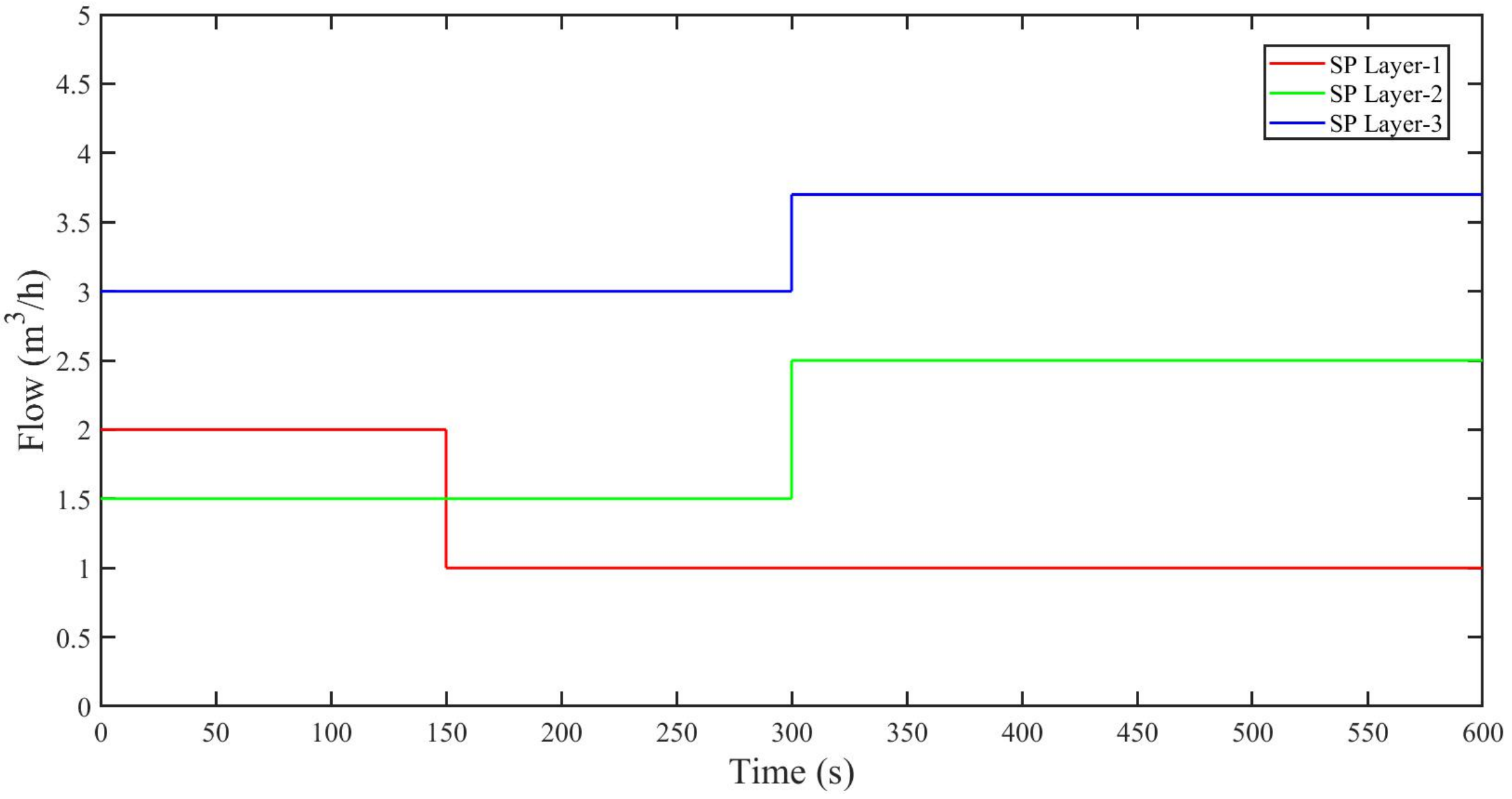

**Fig 10 Flow value setting curves for each layer**

To achieve a rigorous comparison, the control variable principle is used and all control algorithms are selected with the same common parameter metrics to compare the impact of algorithm under independent parameters on the control performance. The non-optimal MPC parameters were selected as shown in the following table2.

**Table 2 Basic MIMO-MPC design parameters**

| Parameters | Value |
|---|---|
| **Control horizon** | 1 |
| **Prediction horizon** | 25 |
| **Error weight ($[Q_1, Q_2, Q_3]$)** | [0.2,0.15,0.5] |
| **Control weight ($[R_1, R_2, R_3]$)** | [5.7,2.5,7] |
| **Correction coefficient ($[h_1, h_2, h_3]$)** | [0.35,0.25,0.15] |

The dynamic performance of the flow value of each layer segment under PID and MPC-PID cascade control algorithms is illustrated in Fig. 11. The output under different algorithms achieve effective tracking of the set value, while the amplitude of the oscillation of the output of the cascade algorithm is significantly weakened, but the steady-state time of following the set value control is basically the same as the result of the PID algorithm, in 450s, 550s the output is subject to The system is more affected under PID control during the interference. The cascade algorithm uses MPC as the main controller, which has better performance than the PID control algorithm in terms of overshoot reduction and disturbance recovery, but the steady-state time is still longer and oscillation exists. the selection of the PID controller parameters is shown in the following table3.

**Table 3 PID controller common parameters**

| Parameters | Value |
|---|---|
| $K_p$ | [2.75, 3.00, 7.67] |
| $T_i$ | [1.00, 0.77, 2.00] |
| $T_d$ | [0.67, 0.37, 1.33] |

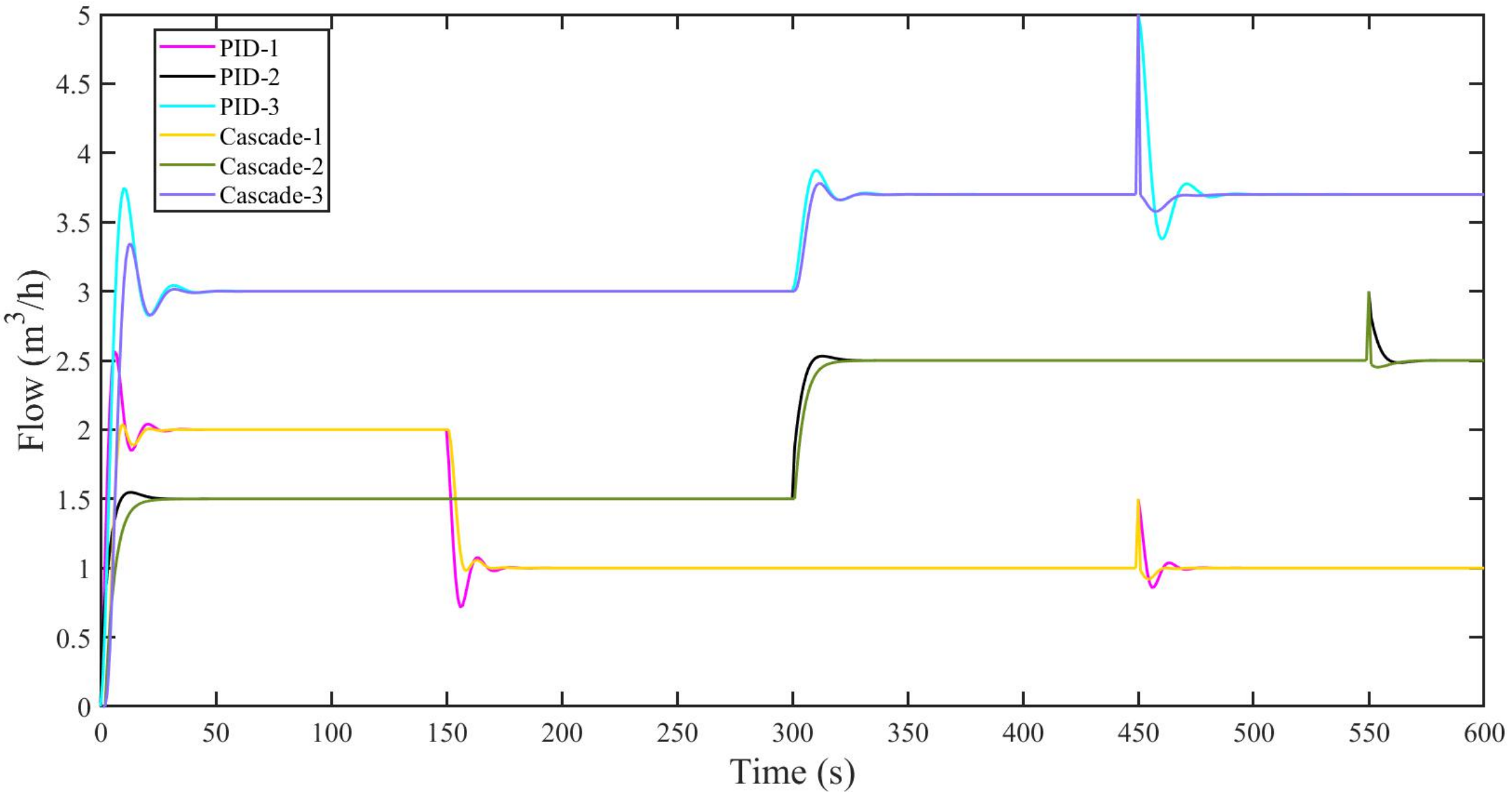

**Fig 11 Layer flow control simulation results of PID and cascade MPC-PID**

Fig. 12 shows the dynamic changes of the flow value in the lower layer segment for MPC, cascaded MPC-PID, and fused three control algorithms. As shown in the figure, all control algorithms eliminate the steady-state error and achieve accurate control of the flow. It is calculated by 3.3 that the control effect of each loop is best when the weight matrix $\boldsymbol{w}$ is taken as $[0.625,0.57,0.25]^T$. The target algorithm maintains the response speed of the string-level control algorithm, which is significantly improved compared with the non-optimized parameter MPC algorithm. It can be seen that the target algorithm does not need to fine-tune the complex parameters of the MPC controller, but only needs to change the weights to complete the process of performance optimization, which simplifies the operation process. Compared with the cascade control algorithm, the amount of oscillation and overshoot is weakened, which is mainly due to the addition of the MPC weight control component and reasonable weight ratios, which shorten the steady-state establishment process time, track the set value faster, and improve the dynamic response performance.

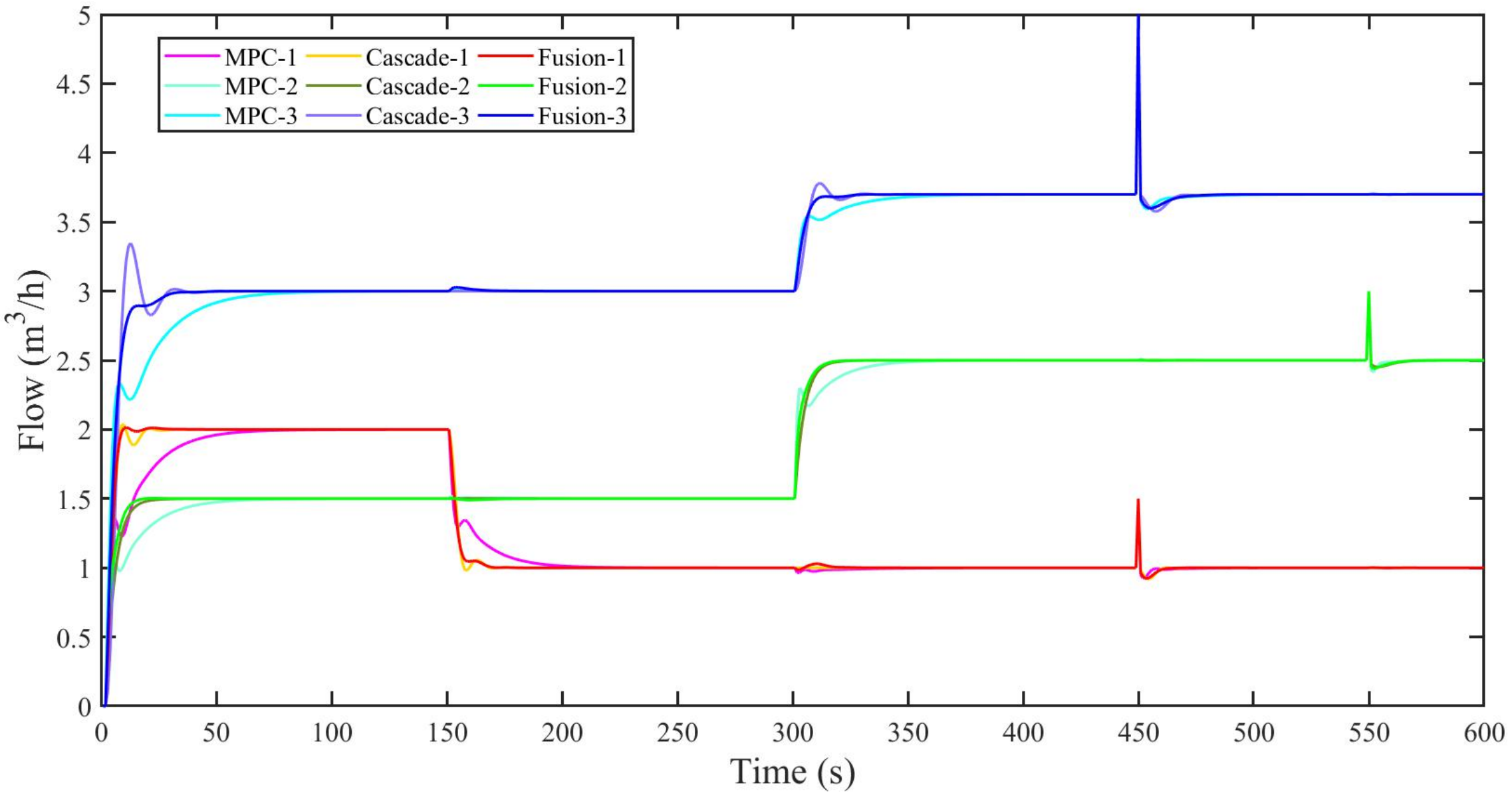

**Fig 12 Layer flow control simulation results of PID, cascade MPC-PID, and fusion**

To verify the dynamic performance of various algorithms comprehensively, several performance metrics were selected for comprehensive comparison, and the results are shown in Table 3. Where $H$ represents the lift of the pump, $\boldsymbol{\sigma_p}$ is the overshoot when the curve starts from 0s and the output reaches the set value, $\boldsymbol{\sigma_d}$ is the overshoot when the output returns to steady state after a strong disturbance at 450s, $\boldsymbol{t_s}$ is the steady state time, and $\sum \boldsymbol{L_u}$ is the cumulative change amplitude of each output control signal during the overall control process.

**Table 3 Dynamic performance index evaluation**

| **Algorithm** | $\boldsymbol{\sigma_p}(\%)$ | | $\boldsymbol{\sigma_d}(\%)$ | | $\boldsymbol{t_s}(\boldsymbol{s})$ | $\sum \boldsymbol{L_u}$ |
|---|---|---|---|---|---|---|
| | $q_v$ | $H$ | $q_v$ | $H$ | | |
| **PID** | 13.65 | 21.49 | 6.45 | 9.23 | 62 | 5.826 |
| **MPC** | 0 | 0 | 2.46 | 6.36 | 99 | 3.826 |
| **Cascade** | 8.15 | 14.29 | 2.54 | 6.80 | 63 | 4.128 |
| **Fusion** | 0 | 0 | 2.42 | 5.81 | 55 | 2.907 |

To test the control performance of each algorithm and the change of the working state of the equipment under the abnormal control state comprehensively, the simulation of the working condition with unstable discrete sampling frequency under the same weight was performed, and the simulation results of the pump head, full well flow, pump efficiency and the control signal curve received by the underground water distributor are shown in Fig. 13.

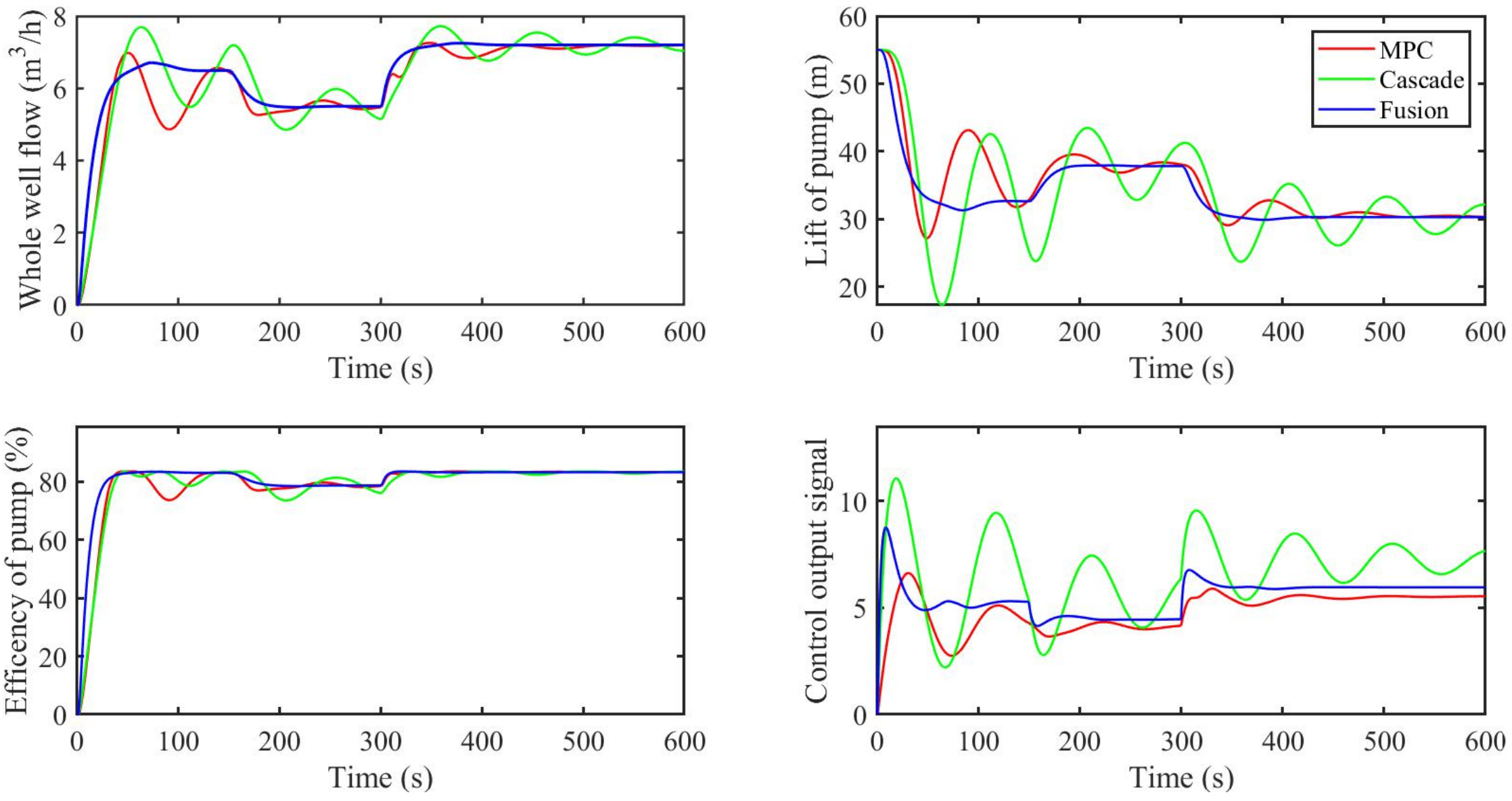

**Fig 13 Variation of equipment operating parameters under poor sampling settings**

In the state of poor sampling settings, the MPC and cascade MPC-PIC algorithm controls the whole well flow and pump efficiency fluctuations, and the pump lift oscillation amplitude is large, while the target algorithm basically maintains the stability of each parameter control and weaken the fluctuation of the water distributor signal. Weakening the fluctuation of full well flow will avoid the water hammer effect and protect the piping equipment while maintaining the stability of efficiency will effectively extend the service life of water distribution equipment.

## 4.2 Wave code communication simulation optimization

Control algorithm optimization is mainly aimed at improving the decoding success rate of intelligent water injection communication. The quality of generation depends on the generation time of the parametric wave code and its approximation to the standard square wave. Fig. 14 shows the actual generated wave codes. In the above figure, the wave code is weak and the wellhead cannot be recognized, resulting in decoding failure. Therefore, the following figure increases the amplitude of openness fluctuation, the wave code pattern is more obvious, and the decoding success rate is improved. However, since the signal curve has a larger error than the standard square wave, there is still a distortion of the decoded data, resulting in the user not being able to collect the real physical data of the well in the upload communication, while the failure of decoding or miscoding of the downhole controller in the downlink communication will lead to mis-operation of the underground water distributor.

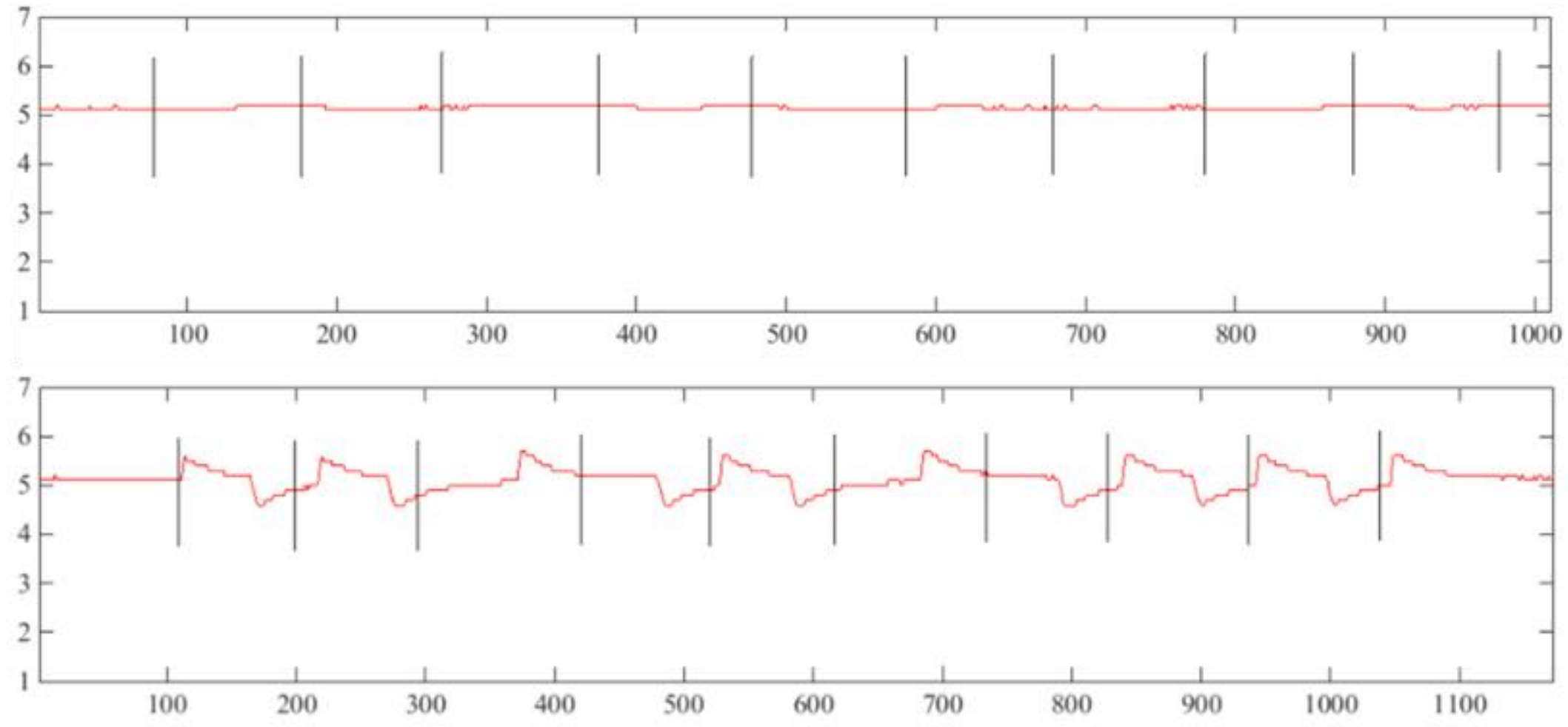

**Fig.14 Actual communication wave code generated by fluid fluctuations**

Fig. 15 illustrates the simulation results of the layer segment flow wave code signal, comparing the wave code effect generated by the three control algorithms. The nozzles in the 1st and 3rd layer change periodically during the 400~1200s stage to generate flow wave coding, in which the 3-layer target signal is 3~3.7m^3/h amplitude, and the first-layer target wave coding amplitude is 1~2m^ 3/h, the later period is 2~2.5m^3/h, 2-layer section maintains 1.5 $m^3/h$ constant. The figure shows that the process of the curve's progressive amplitude in the MPC algorithm is slow, the tracking performance is poor, and the generated wave code has a large curvature. Cascade control algorithm has a certain amplitude of oscillation overshoot, which can destroy the wave code curve structure, resulting in lower decoding pass rate and data distortion. Since the wave code curve of the target fusion control algorithm is closest to the target square wave, it is possible to achieve fast and accurate decoding in this state even with reduced fluctuation amplitude compared to the more stringent wave code amplitude conditions in other algorithms, and to achieve the goal of power-saving by reducing the reciprocal stroke of the valve action.

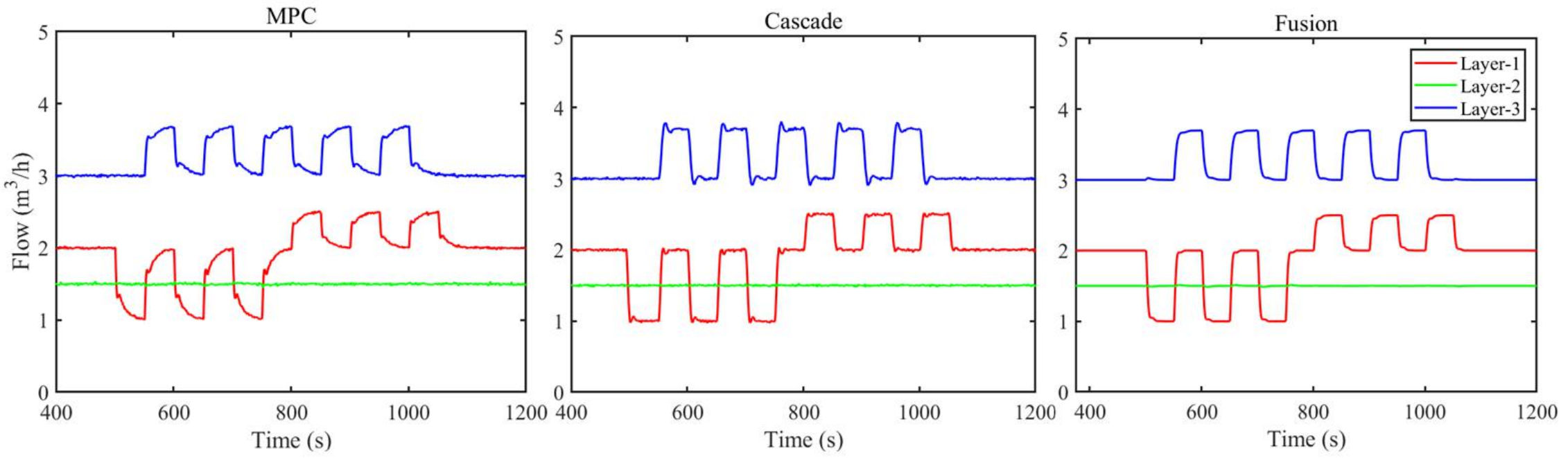

**Fig 15 Simulation results of flow wave code signals for each layer**

For closer approximation to the actual working conditions, this paper studies the control situation when the controller load model and the actual controlled system model are severely mismatched, the system output is more unstable in this state, and the robustness of different control algorithms can be compared critically and clearly without changing the weight matrix. A comparison of the layer section flow wave code signal effects is shown in Fig. 16. It can be seen that the MPC algorithm control recovery is slow, and the wave code amplitude is not fixed, which generates multiple wave peaks and is easy to cause false codes. Compared with the output steady-state of Fig. 15, the defects of the string-level algorithm are more obvious. The amplitude of the output oscillation of the 3-layer segment is further increased compared with the MPC algorithm, and the error is great compared with the control results of the algorithm in Fig. 15, and the controlled signal has a trend of divergence because the underlying PID algorithm amplifies the role of large model mismatch in proportional form and exacerbates the system instability by differentiation, which is more obvious when controlling the model with large steady-state gain. The proposed algorithm maintains the fast response of the rising phase, oscillations are substantially weakened, the wave code signals generated at different periods are of the same amplitude, and the output converges quickly to the original setting after wave code abort. In conclusion, the quality of the generated wave code is relatively high. which is mainly due to the robust effect of MPC superimposed on the attenuating effect of the cascade weighting factor on the mismatch error.

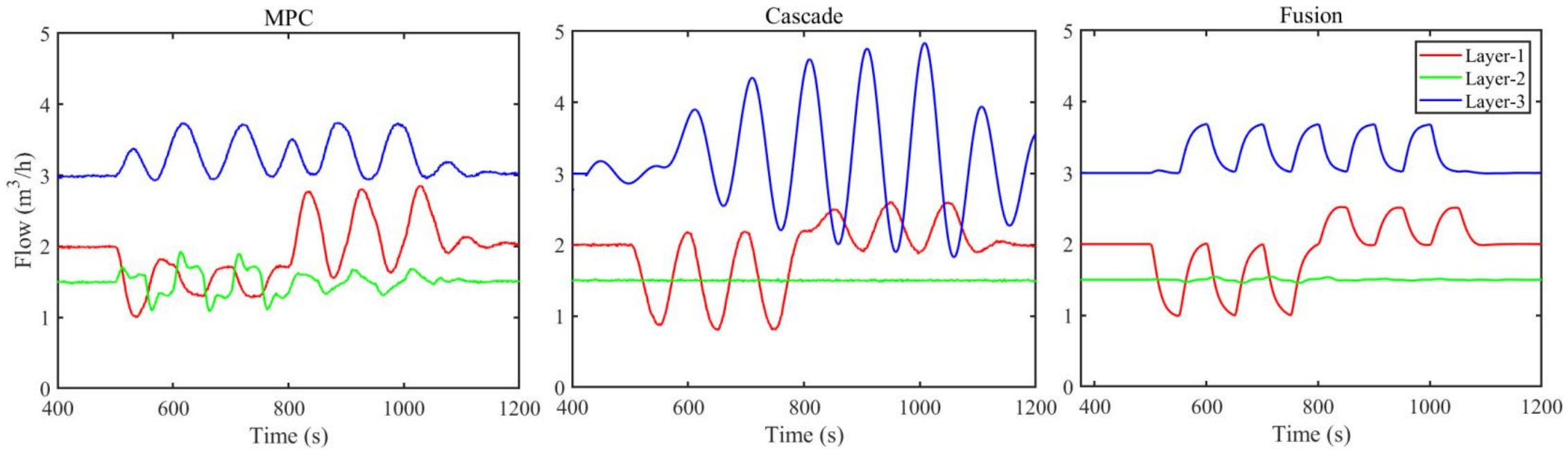

Fig 16 Comparative simulation results of layer segment flow wave code signal robustness

For a more realistic and comprehensive simulation of the controller model mismatch conditions, robust simulations of the full well flow, pressure wave code, water distributor control signal and pumping machine efficiency were performed simultaneously with the wave code generation process. The simulation results are shown in Fig. 17.

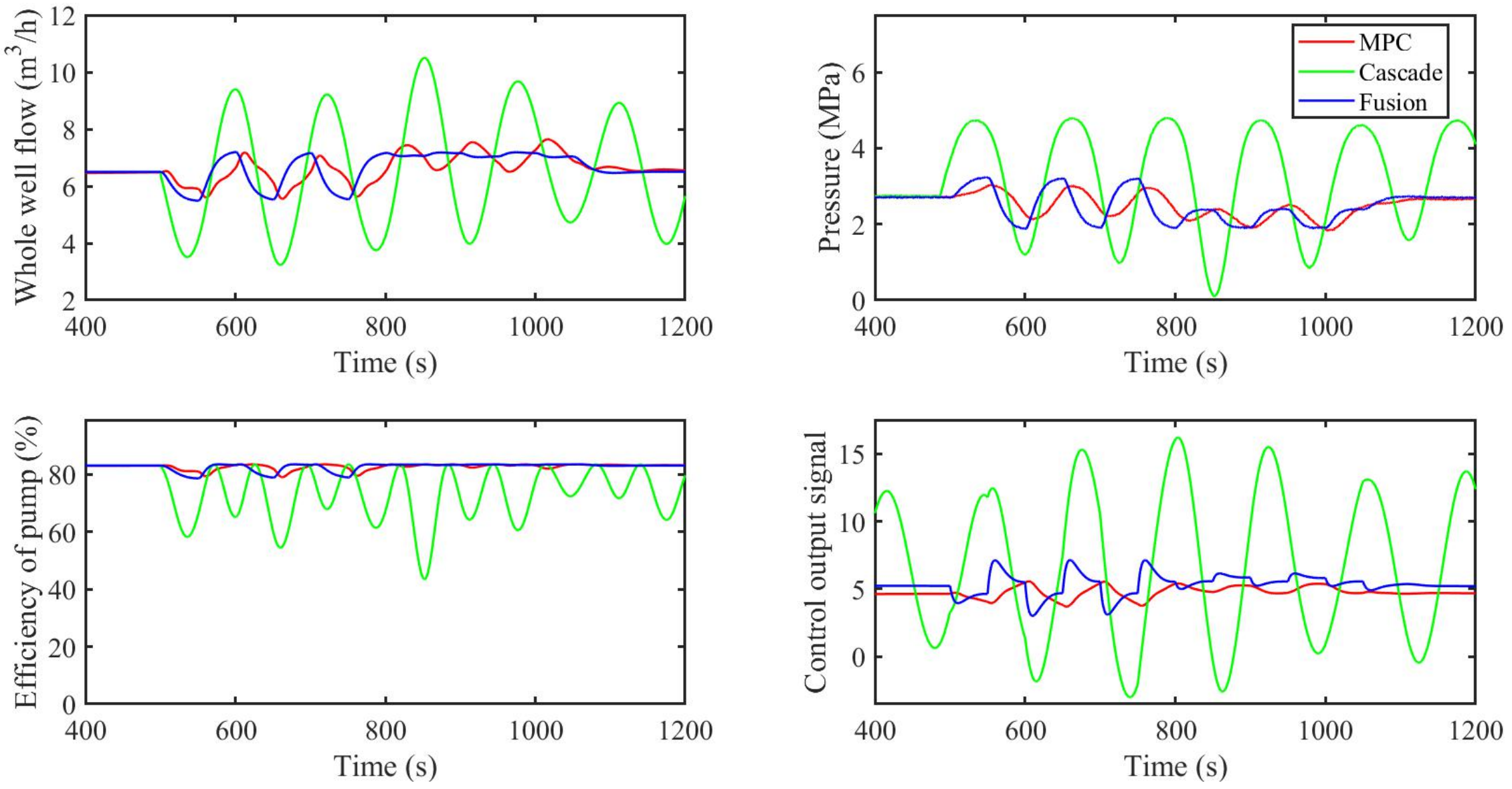

**Fig 17 Comparative simulation results of equipment operating parameters robustness**

As shown in the figure, during the wave code generation phase from 500 to 800s, the wave codes generated in the 1 and 3 layer sections have the same direction, so the whole well flow value changes more under the control of MPC and the proposed algorithm. While the wave codes in the double layer sections have opposite directions from 800 to 1200s, the proposed algorithm maintains a constant whole well flow and avoids ineffective fluctuations that MPC produces, while the control performance of the string-level MPC-PID algorithm deteriorates severely throughout. Meanwhile, the pressure fluctuations under the cascade MPC-PID algorithm are more violent, which may cause sealing device to fail, leading to uncontrolled and undifferentiated water injection in each layer. MPC algorithm generates pressure waves with irregular amplitude and large curvature of wave code signal, which is not conducive to decoding. The fusion control algorithm generates pressure waves of consistent amplitude at different periods, and the waveform is controlled within a reasonable and identifiable range, which is more compliant than other algorithms. MPC-PID cascade algorithm output control signal appears to be overloaded, the output of the negative signal will reverse the voltage, drive the water distributor nozzle repeatedly action, losing power, making energy use efficiency reduced. In the simulation of pumping machine efficiency, the efficiency of the pumping machine under MPC and the proposed algorithm is stable and maintained at about 80%, with less effect from the change of full well flow value. The pumping machine efficiency fluctuates unstably under the cascade algorithm, and the generation of wave code causes the efficiency to drop rapidly to 50%~60%, resulting in larger power loss. While the fusion-controlled pumping machine efficiency is basically maintained at 800s~1200s, which meets the requirements of efficient energy utilization and protection of equipment.

## 4.3 Response characteristics analysis

This section clarifies the mechanism of optimal fusion control by visually demonstrating the fusion effect of MPC and PID algorithms with different weights. Fig. 18 records the output curves for each loop string section weight as the uniform step size increases. The output overshoot of first loop gradually decreases until the output curve fits exactly with the set value, at this point, if the weight value is increased again, it will intensify the vibration of the system and make the steady-state time longer. If the weight in the second output continues to increase beyond the optimal value, it will not increase the overshoot, but the output curve will gradually shift downward away from the set value curve, which will slow down the steady-state establishment process. In the first and second outputs, the optimal weights are selected as the minimum value in the overshoot-free weight set, preserving the favorable performance of the lower-level PID. In the third output, the original cascade algorithm output oscillation is violent and dynamic performance is poor, so reducing the cascade control weights is necessary. The optimal output curve fits exactly with the set value, and at this point, if w continues to be reduced it will weaken the fast response characteristics of the system. Optimal weights can be calculated from (38) to (39).

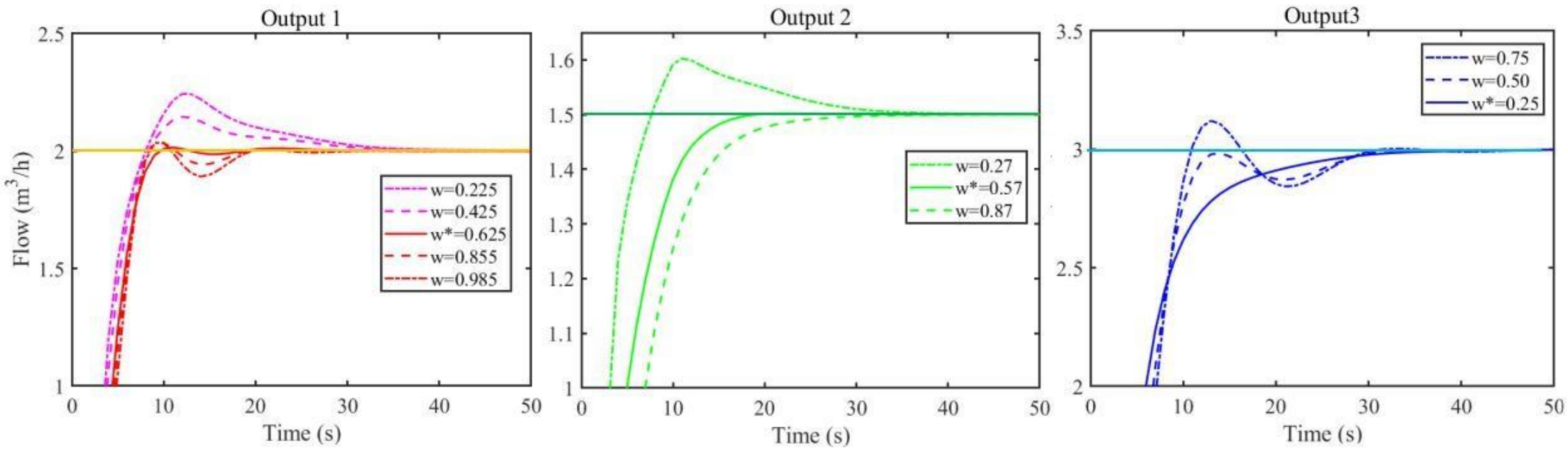

**Fig 18 Output of different weights in different control loops**

To ensure the universality of the optimal dynamic performance of the fusion control, the three output loops cover all types of output curves of the cascade control algorithm: Output 1 overshoot is much smaller, but oscillation exists. Output 2 has no overshoot and no oscillation. Output 3 overshoot and oscillation exist simultaneously. In particular, in the 3-loop, as long as the first peak of the output curve under the specified weights is guaranteed to be no greater than the set value, the overshoot is eliminated, optimizing the control performance of the large steady-state gain object; with good performance of the lower-level PID. All in all, the optimal fusion control has accomplished the task of optimizing the dynamic performance of multivariable inertia link control under the condition that the controller parameter regulation is refined and simplified.

# 5 Conclusion

In conclusion, it is crucial to ensure the speed and reliability of wave code communication between the wellhead and the underground water distributor in intelligent stratified water injection in oil fields. However, the geological characteristics of the reservoir are different in each section of the well, the characteristics of the pipe column and underground water distribution equipment are complex, and there is no effective method to drive the regular fluctuation of the fluid in the pipe column, which makes it impossible to generate high-quality fluid wave code with fast control, so it is difficult to achieve high-efficiency wave code communication. To effectively improve the control response characteristics of fluid wave codes, this paper proposes an MPC-PID based output optimal fusion control method, establishes a steady-state model of layered water injection based on the well structure from the whole well differential pressure-flow relationship, solves the limit amplitude of generated wave codes, and combines the steady-state operating points of different layer sections with the dynamic characteristics of water distribution equipment to complete the model description of the dynamic process of wave code generation. Meanwhile, the internal principle of multivariate MPC control is derived and extended to the fusion control algorithm to solve the optimal fusion control law when minimizing the cost function under fixed weights. For the performance optimization of optimal weights, the fusion mechanism of MPC and PID algorithm is analyzed in depth combining the internal model and frequency domain structure, the property states of the system output under different fusion weights are discussed, and the theoretical solution of the optimal fusion weights is completed. Simulation results show that the fusion control method has fast target value following capability, generates wave codes with high speed and high quality, and ensures the reliability of wave code data under abnormal working conditions. The communication rate can be effectively improved and the operation efficiency of water distribution equipment can be enhanced, saving labor and maintenance costs. The proposed control scheme achieves the goal of intelligent water injection, which is conducive to the improvement of reservoir recovery in complex geological conditions and is significant to the construction of automated and intelligent oil fields.

## Acknowledgments

The author sincerely thanks the support of the PetroChina Exploration and Development Research Institute and the help of senior brothers and sisters.

## REFERENCES

(1) Song, X.M., Qu, D.B., Zou, C.Y., 2021. Low cost development strategy for oilfields in China under low oil prices. J. Petr -oleum Exploration and Development. 48(4), 1007–1018. https://doi.org/10.1016/S1876-3804(21)60085-X.

(2) Li, H.B., Yang, Z.M., Li, R.S., 2021. Mechanism of CO2 enhanced oil recovery in shale reservoirs. In press. J. Petroleum Science. https://doi.org/10.1016/j.petsci.2021.09.040.

(3) Liang, T., Hou, J.R., Qu, M., 2021. Application of nanomaterial for enhanced oil recovery. In press. J. Petroleum Science. https://doi.org/10.1016/j.petsci.2021.11.011.

(4) Mohamed. A., Wu, Z.Y., Zhou, D.Y., 2021. A review of chemical-assisted minimum miscibility pressure reduction in CO2 injection for enhanced oil recovery. Petroleum 7, 245-253. https://doi.org/10.1016/j.petlm.2021.01.001.

(5) Ajoma, E., Saira, Sungkachart T., 2020 Water-saturated CO2 injection to improve oil recovery and CO2 storage. J. Applied Energy 226, 114853. https://doi.org/10.1016/j.apenergy.2020.114853.

(6) Jiang, J., Rui, Z., Hazlett, R., Lu, J., 2019. An integrated technical-economic model for evaluating CO2 enhanced oil recov -ery development. J. Applied Energy 247, 190–211. https://doi.org/10.1016/j.apenergy.2019.04.025.

(7) Wang, X., Klaasvan, V., Marcy, P., 2018. Economic co-optimization of oil recovery and CO2 sequestration. J. Applied Ene -rgy 222, 132-147. https://doi.org/10.1016/j.apenergy.2018.03.166.

(8) Luo, P., Luo, W.G., Li, S., 2017. Effectiveness of miscible and immiscible gas flooding in recovering tight oil from Ba-kk en reservoirs in Saskatchewan, Canada. J. Fuel 208, 626–636. https://doi.org/10.1016/j.fuel.2017.07.044.

(9) Guo, D.L., Kang, Y.W., Wang, Z.Y., 2021. Optimization of fracturing parameters for tight oil production based on genetic algorithm. J. Petroleum, 419, 00089-4. https://doi.org/10.1016/j.petlm.2021.11.006.

(10) Wang, X.L., Wu, P.C., Han, Y.P., 2008. Current situation and measures of water injection in Chang 8 Layer, Xi-feng Oilfield, Changqing Oilfield. J. Petroleum Exploration and Development, 35(3), 344–348.

(11) Zhao, W.B., Hu, S.Y., Deng, X.Q., 2021. Physical property and hydrocarbon enrichment characteristics of tight oil reservoir in Chang 7 division of Yanchang Format-ion, Xin'anbian oilfield, Ordos Basin, China. J. Petroleum Science, 18:1294-1304. https://doi.org/10.1016/j.petsci.2020.07.001.

(12) Zhou, N.W., Lu, S.F., Wang, M., et al., 2021. Limits and grading evaluation criteria of tight oil reservoirs in typical conti-nental basins of China . J. Pet. Sci, 48(5): 1089–1100. https://doi.org/10.1016/S1876-3804(21)60093-9.

(13) Ganat, T.A., Hrairi, M., Farj, M., et al., 2015. Development of a novel method to estimate fluid flow rate in oil wells usi-ng electrical submersible pump. J. Pet.Sci. Eng, 135: 466–475. https://doi.org/10.1016/j.petrol.2015.09.029.

(14) Jing, G.L., Tang, S., Li, X.X., Wang, H.Y., 2017. The analysis of scaling mechanism for water injection pipe columns in t -he Daqing Oilfield Production optimization for alternated separate-layer water injection in complex fault reservoirs. J. Arabi -an Journal of Chemistry 10, 1235–1239. http://dx.doi.org/10.1016/j.arabjc.2013.02.023.

(15) Zhang, L.M., Xu C., Zhang, K., et al., 2020. Production optimization for alternated separate-layer water injection in comple -x fault reservoirs. J. Journal of Petroleum Science and Engineering 193, 107409. https://doi.org/10.1016/j.petrol.2020.107409

(16) Jia, D.L., Liu, H., Zhang, J.Q., et al., 2020. Data-driven optimization for fine water injection in a mature oil field. J. Petro -leum Exploration and Development, 47(3), 674–682. https://doi.org/10.1016/S1876-3804(20)60084-2.

(17) Zhou, Y.P., 2019. Research and application of preset cable intelligent injection technology (in Chinese), J. Oil Field Equipment, 48(4), 69–73. https://doi:10.3969/J.ISSN.1001-3482.2019.04.01.

(18) Li, Z.L., 2019. Development status and application prospect of intelligent separation injection technology (in Chinese). J. New Material a-nd New Technology, 45(7), 1003–6490. 1003﹣6490（2019）07﹣0049﹣02

(19) Tong, Y., 2020. Wireless remote intelligent water injection control system discussion (in Chinese). J. Chemical Engineering & Equipment, 7, 163000. https://doi:10.19566/j.cnki.cn35-1285/tq.2020.07.058.

(20) Yao, B., Yang, L.Z., Yu, J,Z., 2020. Digital layered water injection based on wave code communication (in Chinese). J. China Petroleum Machinery, 48(5), 71–77. https://doi:10.16082/j.cnki.issn.1001－4578.2020.05.012.

(21) Qian, X., Jia, S.K., Huang, K.J., et al., 2021. MPC-PI cascade control for the Kaibel dividing wall column integrated with data-driven soft sensor model. J. Chemical Engineering Science, 231, 116240. https://doi.org/10.1016/j.ces.2020.116240.

(22) Ravendra, S., Marianthi, I., Rohit, R., 2013. System-wide hybrid MPC–PID control of a continuous pharmaceutical tablet manufacturing process via direct compaction. J. European Journal of Pharmaceutics and Biopharmaceutics, 85: 1164–1182. http://dx.doi.org/10.1016/j.ejpb.2013.02.019.

(23) Ravendra, S., Abhishek, S., Krizia, M., et al., 2014. Implementation of an advanced hybrid MPC–PID control system using PAT tools into a direct compaction continuous pharmaceutical tablet manufacturing pilot plant. J. International Journal of Pharmaceutics, 473: 38–54. https://doi.org/10.1016/j.ces.2020.116240.

(24) NI, Y.M., 2013. Fusion Control Strategy Based on I-Fuzzy-Smith Algorithm for Complex Process with Large Lag. J. Hydro Mechatronics Engineering, 41(18): 1001–3881. https://doi:10.969/j.issn.1001－3881.2013.18.031.

(25) Deng, Z.H., Chen, Q.H., Zhang, L.Y., et al., 2020. Weighted fusion control for proton exchange membrane fuel cell system. J. International Journal of Hydrogen Energy, 45: 15327–15355. https://doi.org/10.1016/j.ijhydene.2020.01.137.

(26) Severiano, J., Silva, A., Sussush, A., et al., 2019. Corrosion damages off low regulation valves for water injection in oilfiel-ds. J. Engineering Failure Analysis, 96: 362–373. https://doi.org/10.1016/j.engfailanal.2018.11.002.

(27) Gao, S.S., Yi, Y., Liao, G.Z., 2022. Experimental research on inter-fracture asynchronous injection-production cycle for a horizontal well in a tight oil reservoir. J. Journal of Petroleum Science and Engineering, 208: 109647. https://doi.org/10.1016/j.petrol.2021.109647.

(28) Han, B., Cui, G.D., Wang, Y.Q., et al., 2021. Effect of fracture network on water injection huff-puff for volume stimulation horizontal wells in tight oil reservoir: Field test and numerical simulation study, 207:109106. https://doi.org/10.1016/j.petrol.2021.109106.

(29) Zhang, X., Yang, S.L., Wen, B., et al., 2013. Experimental study on threshold pressure gradient of CO2 miscible flooding in low permeability reservoir, 35(5): 1001-6112. https://doi:10.11781/sysydz201305583.

(30) Xiong, W., Lei, Q., Liu X.G., et al., 2009. Pseudo threshold pressure gradient to flow for low permeability reservoirs. J. Petroleum Exploration and Development, 36(2): 1000-0747. 1000-0747(2009)02-0232-05

(31) Jiang, T.W., Huang, Z.W., Li, J.B., Zhou, Y.S., 2021. Internalflow mechanism of cone-straight nozzle. J. Petroleum Science, 18:1507-1519. https://doi.org/10.1016/j.petsci.2021.08.008.

(32) Eugenio, A., Vicenç, P., Joseba, Q., 2019. LPV-MPC Control for Autonomous Vehicles. J. IFAC PapersOnLine, 52-28: 106–113. https://10.1016/j.ifacol.2019.12.356.